\documentclass[epj]{svjour}
\usepackage{graphicx}
\usepackage{amsmath,amssymb}
\usepackage{bm}

\begin{document}
\title{A still unsettled issue in the nucleon spin decomposition problem : 
On the role of surface terms and gluon topology}
\author{Masashi Wakamatsu\inst{1}
}                     
\offprints{}          
%
\institute{KEK Theory Center, Institute of Particle and Nuclear Studies,\\
High Energy Accelerator Research Organization (KEK),\\
1-1, Oho, Tsukuba, Ibaraki 305-0801, Japan}
\date{Received: date / Revised version: date}
%
\abstract{In almost all the past analyses of the decomposition of the nucleon spin into its constituents,
surface terms are simply assumed to vanish and not to affect the integrated sum rule of
the nucleon spin. However, several authors claim that neglect of surface terms is not necessarily
justified, especially owing to possible nontrivial topological configuration of the gluon field in the
QCD vacuum. There also exist some arguments indicating that the nontrivial gluon topology
would bring about a delta-function type singularity at zero Bjorken variable into the longitudinally
polarized gluon distribution function, thereby invalidating a naive partonic sum rule for the total
nucleon spin. In the present paper, we carefully examine the role of surface terms in the nucleon spin
decomposition problem. We shall argue that surface terms do not prevent us from obtaining
a physically meaningful decomposition of the nucleon spin. In particular, we demonstrate that
nontrivial topology of the gluon field would not bring about a delta-function type singularity into
the longitudinally polarized gluon distribution functions. We also make some critical comments on the
recent analyses of the role of surface terms in the density level decomposition of the total
nucleon angular momentum as well as that of the total photon angular momentum.
} 

\titlerunning{The role of surface terms and gluon topology in the nucleon spin decomposition problem}
\authorrunning{M.~Wakamatsu}

\maketitle
\section{Introduction}
\label{intro}

How to decompose the net spin of the nucleon into its intrinsic spin and
orbital angular momentum parts of quarks and gluons is one of the 
fundamental problem of quantum chromodynamics (QCD). 
(See \cite{LL2014},\cite{Waka2014} for reviews.)
Popular for a long time were the two decompositions known as
the Jaffe-Manohar decomposition \cite{JM1990} and 
the Ji-decomposition \cite{Ji1997}, both of
which have their own merits and demerits. The Jaffe-Manohar
decomposition offers a complete decomposition of the nucleon spin
into the spin and orbital parts of quarks and gluons, but it is not a
gauge-invariant decomposition. On the other hand, the Ji decomposition
is a manifestly gauge-invariant decomposition at the cost of giving
up the decomposition of the total gluon angular momentum into
its spin and orbital parts. 
After long and intense debates (see, for example, \cite{LL2014},\cite{Waka2014}
for review), many researchers now believe that the Jaffe-Manohar 
decomposition can be made 
gauge-invariant at least formally by introducing the concept of the physical 
component of the gauge field first introduced into the problem by 
Chen et al. \cite{Chen2008},\cite{Chen2009}. 
However, if one is allowed to use the
idea of the physical component of the gauge field, the total gluon
angular momentum can also be decomposed gauge-invariantly into
its spin and orbital parts. As a consequence, we are led to
two complete gauge-invariant decomposition of the nucleon spin. 
The one is the improved Jaffe-Manohar decomposition \cite{BJ1998}, 
which belongs to the decomposition of canonical type, and the other is 
an extension of the Ji-decomposition, which we call the decomposition of
mechanical (or kinetic) type \cite{Waka2010},\cite{Waka2011}. 
(We recall, however, that
these two types of decomposition has totally different physical contents.
What reflects the intrinsic spin structure of the nucleon is the
mechanical type decomposition not the canonical type one,
as throughly explained in the recent papers \cite{Waka2015},\cite{Waka2016}. 
See also \cite{Waka2013}.)

It is important to recognize the fact that
different decompositions mentioned above are related 
to each other by the addition or subtraction of surface terms.
Little attention has been paid to these surface terms, since
the surface terms were believed not to contribute 
to the integrated sum rule of the nucleon spin.
However, several authors suspect that the surface terms may not
necessarily be neglected and rather they may play some unexpected roles
in the nucleon spin decomposition 
problem \cite{Lowdon2014}-\nocite{Bass2005}\nocite{Bass2009}
\nocite{Tiwari2015}\nocite{Nayak2018A}\cite{Nayak2018B}. 
For example, based on
a rigorous field theoretical treatment of the forward nucleon matrix
elements of the relevant two surface terms of the angular momentum
tensor, Lowdon concluded that
the forward nucleon matrix elements of these surface terms precisely
cancel the corresponding matrix elements for the quark and gluon spin 
terms \cite{Lowdon2014}.
This is a perplexing conclusion, since, if it were true, the quark and
gluon spin terms do not contribute to the net nucleon spin sum rule
after all, thereby bereaving the practical significance of the sum rule.

Other authors claim that the nontrivial topological configuration 
of the gluon field in the QCD vacuum may play an unforeseen role also 
in the nucleon spin decomposition 
problem \cite{Bass2005}-\nocite{Bass2009}\nocite{Tiwari2015}
\nocite{Nayak2018A}\cite{Nayak2018B}. Among others, Bass insisted
that the nontrivial gluon topology would bring about a delta-function
type singularity at zero Bjorken variable ($x = 0$) into the longitudinally
polarized gluon distribution function \cite{Bass2005},\cite{Bass2009}. 
We recall that the existence of such a delta-function type singularity
is already known for several quark distribution functions.
For example, within the framework of perturbative QCD at the
one-loop level, Burkardt and Koike suggested the existence of the delta-function
singularity at $x = 0$ in the twist-3 distribution functions $h_L (x)$ and
$e (x)$ \cite{BK2002}. (See also more recent analysis \cite{AB2018}.)
The existence of the delta-function singularity in the twist-3 unpolarized
distribution $e (x)$ was also confirmed not only within the framework
of perturbative QCD but also within the framework of nonperturbative 
QCD \cite{WO2003},\cite{ES2003}.
It was confirmed also within the framework of an effective model
of the nucleon, i.e. the chiral quark soliton model, based on which explicit
numerical calculation of $e (x)$ was also carried 
out \cite{OW2004},\cite{Schweitzer2003}.
This fact has an important phenomenological impact in the respect 
that the existence of the delta-function singularity in $e (x)$ means
breakdown of the pion-nucleon sigma term sum rule
$\int_{- \,1}^1 \,e(x) \,d x = \bar{\sigma} = \Sigma_{\pi N} / \bar{m}$ for
the experimentally accessible distribution function $e (x)$, where $\bar{\sigma}$ 
is the scalar charge of the nucleon, while $\Sigma_{\pi N}$ is the pion-nucleon 
sigma term with $\bar{m}$ being the average of the up- and down-quark masses.
Similarly, if the longitudinally polarized gluon distribution has a delta-function
type singularity, it has a danger of bereaving the partonic sum rule
for net nucleon spin \cite{Bass2005},\cite{Bass2009}. 
It is therefore of vital importance to carefully
check whether such a singularity in the longitudinally polarized gluon
distribution really exists or not.

In addition to the integrated sum rule of the nucleon spin, there also exists
interest in the density level decomposition of the nucleon total angular
momentum as well as in the density level decomposition of the total photon 
angular momentum.
If one is interested in the density level decomposition of the total angular
momentum, one has no reason to drop these surface terms.  
In recent papers \cite{Leader2016},\cite{Leader2018}, Leader investigated 
the density level decomposition of the 
total angular momentum of a free photon beam into orbital and spin parts.
On the other hand, the density level decomposition of the total
angular momentum of the nucleon was addressed by Lorc\'{e}, Mantovani, and
Pasquini \cite{LMP2018}. 
Both authors emphasized that, at the density level, an unguarded neglect
of the surface terms can never be justified, because the difference of 
two different decompositions of the total angular momentum are just
characterized by these surface terms.

Now the purpose of the present paper is to answer all the questions raised 
above as convincingly as possible.
The paper is organized as follows. First, in sect.2,
we carefully inspect Lowdon's proof that the forward nucleon matrix
element of the relevant surface terms precisely cancel the corresponding
matrix elements for the quark and gluon spin terms. Next, in sect.3,
we try to ask the question whether the nontrivial gluon topology really
brings about the delta-function type singularity in the longitudinally
polarized gluon distributions. We argue that the answer to this question
critically depends on the rigorous definition of the longitudinally polarized gluon
distribution, which has been left in unclear status for a long time. 
In sect.4, we briefly review
the essence of Leader's paper as well as Lorc\'{e} et al.'s paper on the
density level decomposition of the total angular momentum of the free photon
and that of the nucleon from our own perspective. After that, we make several
critical comments on their analyses and main claims. 
Then, in sect.5, we summarize principle conclusions drawn from the
present investigation.

\section{On Lowdon's quantum field theoretical analysis of boundary terms}
\label{sec:2}

Lowdon's argument starts with the following expression for the QCD
angular momentum tensor \cite{Lowdon2014}, which was first written down 
in the paper by Jaffe and Monohar  \cite{JM1990}: 
\begin{eqnarray}
 M^{\mu \nu \lambda}_{QCD} &=& 
 \frac{i}{2} \,\bar{\psi} \,\gamma^\mu \,
 (x^\nu \,\partial^\lambda - x^\lambda \,\partial^\nu) \,\psi \ + \ \mbox{h.c.}
 \nonumber \\
 &+& \frac{1}{2} \,\epsilon^{\mu \nu \lambda \rho} \,
 \bar{\psi} \,\gamma_\rho \,\gamma^5 \,\psi \nonumber \ \ \ \\
 &-& F^{\mu \rho a} \,(x^\nu \,\partial^\lambda - x^\lambda \,\partial^\nu) \,
 A^a_\rho \nonumber \\ 
 &+& F^{\mu \lambda a} \,A^{\nu a} 
 \ + \ F^{\nu \mu a} \,A^{\lambda a}
 \nonumber \\
 &+& \frac{1}{4} \,F^a_{\alpha \beta} \,F^{\alpha \beta a} \,
 (x^\nu \,g^{\mu \lambda} - x^\lambda \,g^{\mu \nu}) \nonumber \\
 &-& \frac{i}{16} \,\partial_\beta \left[ x^\nu \,\bar{\psi} \,
 \left\{ \gamma^\lambda, [ \gamma^\mu, \gamma^\beta] \right\} \,\psi
 \ - \ (\nu \leftrightarrow \lambda) \right] \ \ \ \ \ \nonumber \\ 
 &+& \partial_\beta \,\left( x^\nu \,F^{\mu \beta a} \,A^{\lambda a}
 \ - \ x^\lambda \,F^{\mu \beta a} \,A^{\nu a} \right),
\end{eqnarray} 
with $a$ being the color index.
If one drops the last two surface terms, the angular momentum tensor reduces to
the form : 
\begin{eqnarray}
 M^{\mu \nu \lambda}_{QCD} &=& 
 \frac{i}{2} \,\bar{\psi} \,\gamma^\mu \,
 (x^\nu \,\partial^\lambda - x^\lambda \,\partial^\nu) \,\psi \ + \ \mbox{h.c.}
 \nonumber \\
 &+& \frac{1}{2} \,\epsilon^{\mu \nu \lambda \rho} \,
 \bar{\psi} \,\gamma_\rho \,\gamma^5 \,\psi \nonumber \\
 &-& F^{\mu \rho a} \,(x^\nu \,\partial^\lambda - x^\lambda \,\partial^\nu) \,
 A^a_\rho \nonumber \\
 &+& F^{\mu \lambda a} \,A^{\nu a} \ + \ F^{\nu \mu a} \,A^{\lambda a}
 \nonumber \\
 &+& \frac{1}{4} \,F^a_{\alpha \beta} \,F^{\alpha \beta a} \,
 (x^\nu \,g^{\mu \lambda} - x^\lambda \,g^{\mu \nu}) ,
\end{eqnarray} 
which provides us with the basis of the famous Jaffe-Manohar decomposition of the
nucleon spin \cite{JM1990}. 
Lowdon suspects the validity of this operation, and examined
the advisability of neglecting the surface terms within a rigorous 
framework of quantum field theory. To this end, he first starts with the angular momentum
charge given by
\begin{equation}
 J^i_{QCD} \ \equiv \ \frac{1}{2} \,\epsilon^{i j k} \,\int d^3 x \,M^{0 j k}_{QCD} (x) .
\end{equation}
To be more rigorous, the angular momentum charges corresponding to 
deep-inelastic-scattering (DIS) observables, i.e. the first moments of the relevant parton
distribution functions (PDFs), are related to the $+$-component in the 
light-cone coordinate not the time-component of the angular momentum tensor, i.e.
\begin{equation}
 J^i_{QCD} \ \equiv \ \frac{1}{2} \,\epsilon^{i j k} \,\int d^3 x \,M^{+ j k}_{QCD} (x) .
\end{equation}
However, this difference is not of vital  importance in our discussion below. We therefore 
follow Lowdon's original expression, for simplicity.

Retaining the surface terms, he is then led to the decomposition : 
\begin{equation}
 J^i_{QCD} \ = \ L^i_q \ + \ S^i_q \ + \ L^i_G \ + \ S^i_G \ + \ S^i_1 \ + \ S^i_2 ,
\end{equation}
where
\begin{eqnarray}
 L^i_q &=& \epsilon^{i j k} \,\int \,d^3 x \,
 \left[\, \frac{i}{2} \,\bar{\psi} \,\gamma^0 \,(x^j \,\partial^k) \,\psi \ + \ \mbox{h.c.} \right], \\
 S^i_q &=& \epsilon^{i j k} \,\int \,d^3 x \,\left[\, \frac{1}{4} \,\epsilon^{0 j k l} \,
 \bar{\psi} \,\gamma_l \,\gamma^5 \,\psi \right], \\
 L^i_G &=& - \,\epsilon^{i j k} \, \int \,d^3 x \,\left[\, F^{0 l a} \,(x^j \,\partial^k) \,A^a_l \,\right], \\
 S^i_G &=& \epsilon^{i j k} \,\int \,d^3 x \,\left[\, F^{0 k a} \,A^{j a} \,\right], \\
 S^i_1 &=& - \,\frac{i}{16} \,\epsilon^{i j k} \,\int \,d^3 x \,\,\partial^l 
 \left[ x^j \,\bar{\psi} \left\{ \gamma^k, [\gamma^0, \gamma^l] \right\} \psi \,\right], 
 \ \ \ \ \\
 S^i_2 &=& \epsilon^{i j k} \,\int \,d^3 x \,\,\partial_l 
 \left( x^j \,F^{0 l a} \,A^{k a} \right) .
\end{eqnarray}
Here, the first four terms correspond to the quark orbital angular momentum (OAM) term,
the quark spin term, the gluon OAM term, and the gluon spin term, respectively, whereas
the last two represent the surface integral terms that may vanish or may not vanish.
Lowdon demonstrated that the forward nucleon matrix elements of the two surface terms
do not vanish and rather that they precisely cancel the corresponding matrix
elements for the quark and gluon spin terms in such a way that
\begin{eqnarray}
 \langle p, s \,|\, S^i_1 \,|\,p, s \rangle \ = \ - \,\langle p, s \,|\, S^i_q \,|\, p, s \rangle, \\
 \langle p, s \,|\, S^i_2 \,|\,p, s \rangle \ = \ - \,\langle p, s \,|\, S^i_G \,|\, p, s \rangle ,
\end{eqnarray}
where $|\,p, s \rangle$ stands for the nucleon state with momentum $p_\mu$ and spin $s_\mu$.
If this conclusion of his were true, the quark and gluon spin terms do not contribute to the
nucleon spin sum rule after all, so that the physical interpretation of the sum rule would be
totally lost.

In view of serious impact of Lowdon's conclusion on the nucleon spin decomposition
problem, an immediate question is
whether there is any oversight in his argument. One should remember that
the problem of surface terms is nothing specific to the nucleon spin decomposition 
problem. Naturally, it also appears in the decomposition
of the total photon angular momentum into its spin and orbital parts.
Worthy of special mention here is the fact that the analogous cancellation between 
the surface integral term and the photon spin term happens also in the photon angular
momentum decomposition problem. In fact, as argued by 
Stewart \cite{Stewart2005}, the total photon angular 
momentum of a free photon can also be decomposed into three pieces, i.e. the  
OAM part, the spin part, and the surface integral term.
He showed that, if one considers a plane wave of arbitrary polarization,
the surface integral term does not vanish and it precisely cancels the photon spin term.
Note that this is just analogous to Lowdon's observation in the nucleon spin case.
According to Stewart, however, this is due to fairly singular nature of a plane 
wave that has infinite spatial extension. Realistic experiments are always carried
out on beams that are of finite extent constricted by some apparatus.
For such realistic photon beam, the surface integral term vanishes after
spatial integration, and the decomposition of the total photon angular momentum 
into its spin and orbital parts has a proper interpretation. 

Coming back to the nucleon spin decomposition problem, we recall that the original proof 
that the forward nucleon matrix element of the surface
integral terms vanish was given by Jaffe and Manohar by using a plane wave nucleon
state $|\,p, s \rangle$ with momentum $p_\mu$ and spin $s_\mu$ \cite{JM1990}.
As emphasized by them, an oversimple analysis of the QCD angular 
momentum tensor using a plane wave state
would easily lead to erroneous conclusions. The reason is that the forward limit
of the nucleon matrix element and the spatial integral do not commute.
To avoid this delicate nature of using a plane wave state, several authors advocated
to use a wave packet state that has finite spatial extension \cite{SW2000},\cite{BLT2004}. 
Lowdon's treatment is thought to be a field-theoretically
more rigorous make-up of the wave packet formalism. 
In this sense, his conclusion that the forward matrix element of the surface integral
terms do not vanish sounds a little strange to us, especially in view of our comment
above on the photon spin decomposition problem.
If any, where is an oversight in Lowdon's argument? 
To answer this question, we need to follow more closely the core of his demonstration.

According to Lowdon, in classical theory, charges are defined as spatial integral of
the time-component of some current density $j^k (x)$ :
\begin{equation}
 Q \ = \ \int \,d^3 x \,j^0 (x). 
\end{equation}
In quantum field theory, however, more rigorous definition of charges should be given by
\begin{equation}
 Q \ = \ \int \,d^4 x \, f(x) \,j^0 (x) \ \equiv \ j^0 (f) ,
\end{equation}
with use of some space-time test function $f$, which works to incorporate the space-time
localization of physical states. A convenient choice of the test function can,
for example, be given by
\begin{equation}
 f (x) \ = \ \alpha (x_0) \,f_R (\bm{x}) ,
\end{equation}
with real functions 
$\alpha \in {\cal D} (\mathbb{R}) \,(\mbox{supp} (\alpha) \subset [- \,\delta, \delta], \delta > 0)$
and $f_R \in {\cal D} (\mathbb{R}^3)$ satisfying
\begin{equation}
 \int d x_0 \,\alpha (x_0) = 1, \hspace{1mm}
 f_R (\bm{x}) = \left\{ \begin{array}{ll}
 1, \ & \ |\bm{x}| \ < \ R \\
 0, \ & \ |\bm{x}| \ > \ R \,(1 + \epsilon), \\
 \end{array} \right.
\end{equation}
with some large enough radius $R$ and with $\epsilon > 0$. 
On the basis of this setting, he derived a crucial relation given below for the surface
integral term (or the super-potential operator)
$\int \,d^3 x \,\,\partial_i \left( x^j \,B^{k 0 i} (x) \right)$. It is given as 
\begin{eqnarray}
 &\,& \langle p \,| \int d^3 x \,\,\partial_i \left( x^j \,B^{k 0 i} (x) \right) |\, 0 \rangle
 \nonumber \\
 &=&\! \left\{ \begin{array}{l}
 \ \lim_{R \rightarrow \infty} \,\int d^3 x \,f_R (\bm{x}) \,
 \langle 0 \,|\, B^{k 0 j} (0) \,|\, 0 \rangle \\
 \hspace{55mm} \mbox{for} \ p = 0 ,\\
 \ \lim_{R \rightarrow \infty} \,\int d^4 x \,\alpha (x_0) \,
 f_R (\bm{x}) \,e^{\,i \,p_\mu \,x^\mu} \\
 \hspace{4mm} \times \left[ \langle p \,|\, B^{k 0 j} (0) \,|\, 0 \rangle \, + \ 
 i \,p_i \,\langle p \,|\, x^j \,B^{k 0 i} (0) \,|\,0 \rangle \right] \\
 \hspace{55mm}  \mbox{for} \ p \neq 0, \\
 \end{array} \right. \ \ \ \ \ \label{Eq:Basic_Lowdon}
\end{eqnarray}
where $|\,p \rangle$ is some momentum eigenstate of the nucleon. 
Since this is the central formula leading to his remarkable conclusion, i.e. the
cancellation between the surface integral terms and the quark and gluon spin
terms, let us reexamine its derivation with extreme care.

We first note that the l.h.s. of Eq.(\ref{Eq:Basic_Lowdon}) can be rewritten as
\begin{eqnarray}
 S &\equiv& \langle p \,|\, \int d^3 x \,\,\partial_i \left( x^j \,B^{k 0 i} (x) \right) |\, 0 \rangle
 \nonumber \\
 &=& \langle p \,| \int d^3 x \,\,\partial_i \left( x^j \,e^{\,i \,\hat{P}_\mu \,x^\mu} 
 B^{k 0 i} (0) \,e^{\,- \,i \,\hat{P}_\mu \,x^\mu} \right) |\, 0 \rangle \nonumber \\
 &=& S_1 \ + \ S_2 \ + \ S_3,
\end{eqnarray}
where
\begin{eqnarray}
 S_1 &=& \langle p \,| \int \! d^3 x \,\,x^j \,i \,\hat{P}_i \,
 e^{\,i \,\hat{P}_\mu \,x^\mu} B^{k 0 i} (0) \,
 e^{\,- \,i \,\hat{P}_\mu \,x^\mu} \,|\, 0 \rangle , \\
 S_2 &=& \langle p \,| \int \! d^3 x \,\,x^j \,
 e^{\,i \,\hat{P}_\mu \,x^\mu} B^{k 0 i} (0) \,
 e^{\,- \,i \,\hat{P}_\mu \,x^\mu}
 ( - \,i \hat{P}_i ) |\, 0 \rangle , \ \ \ \ \ \\
 S_3 &=& \langle p \,| \int \! d^3 x \,\,\delta_{i j} \,
 e^{\,i \,\hat{P}_\mu \,x^\mu} B^{k 0 i} (0) \,
 e^{\,- \,i \,\hat{P}_\mu \,x^\mu} \,|\, 0 \rangle , 
\end{eqnarray}
with $\hat{P}_\mu$ being momentum operator.
Using the relations $\hat{P}_\mu \,|\,p \rangle = p_\mu \,|\,p \rangle$ and
$\hat{P}_\mu \,|\,0 \rangle = 0$, we immediately find that
\begin{eqnarray}
 S_2 \ &=& \ \hspace{1mm} 0, \\
 S_3 \ &=& \ \int \,d^3 x \, \,e^{i \,p_\mu \,x^\mu} \,
 \langle p \,|\, B^{k 0 j} (0) \,|\, 0 \rangle . \label{Eq:S3}
\end{eqnarray}
Somewhat delicate is the first term $S_1$. Here, care must be paid to the non-commuting
nature of the coordinate and momentum operator, i.e.
\begin{equation}
 \left[ x^j, \,\hat{P}_i \right] \ = \ i \,\delta^j_i . \label{Eq:C.R.}
\end{equation}
Taking care of this cation, the $S_1$ term can be rewritten as
\begin{eqnarray}
 S_1 &=& \langle p \,| \int d^3 x \left[ i \,p_i \,x^j \,
 e^{\,i \,\hat{P}_\mu \,x^\mu} \,B^{k 0 i} (0) \,
 e^{\,- \,i \,\hat{P}_\mu \,x^\mu} \,|\, 0 \rangle \right. \ \ \ \nonumber \\
 &\,& \hspace{7mm} \left. \ + \ i^2 \,\delta^j_i \,
 e^{\,i \,\hat{P}_\mu \,x^\mu} \,B^{k 0 i} (0) \,
 e^{\,- \,i \,\hat{P}_\mu \,x^\mu} \,|\, 0 \rangle \right] |\,0 \rangle \nonumber \\
 &=& \int \,d^3 x \,\,e^{\,i \,p_\mu \,x^\mu} \,\,i \,p_i \,
 \langle p \,|\, x^j \,B^{k 0 j} (0) \,|\,0 \rangle \nonumber \\ 
 &\,& \hspace{10mm} \ - \ \int \,d^3 x \,e^{\,i \,p_\mu \,x^\mu} \,
 \langle p \,|\, B^{k 0 j} (0) \,|\, 0 \rangle . \ \ \ \ \ 
\end{eqnarray}
We notice that the second term of the above equation precisely cancels the $S_3$ 
term given by (\ref{Eq:S3}). As a consequence, we eventually get
\begin{eqnarray}
 S &=& \ S_1 \ + \ S_2 \ + \ S_3 \nonumber \\
 &=& \int \,d^3 x \,e^{\,i \,p_\mu \,x^\mu} \,i \,p_i \,
 \langle p \,|\, x^j \,B^{k 0 i} (0) \,|\, 0 \rangle . \ \ \ \ \ 
\end{eqnarray}
This especially means that, for $p = 0$, it holds that
\begin{equation}
 S \ = \ 0 .
\end{equation}
On the other hand, for $p \neq 0$, we obtain
\begin{eqnarray}
 S &=& \int \,d^3 x \,e^{\,i \,p_\mu \,x^\mu} \,i \,p_i \,
 \langle p \,|\, x^j \,B^{k 0 i} (0) \,|\, 0 \rangle \nonumber \\
 &=& \lim_{R \rightarrow \infty} \,\int \,d^4 x \,\alpha (x_0) \,f_R (\bm{x}) \,
 e^{\,i \,p_\mu \,x^\mu} \nonumber \\
 &\,& \hspace{20mm} \times \ i \,p_i \,
 \langle p \,|\, x^j \,B^{k 0 i} (0) \,|\, 0 \rangle . \ \ \ \ \ 
\end{eqnarray}
In the end, our answer can be summarized as
\begin{eqnarray}
 &\,& \langle p \,|\, \int \,d^3 x \,\,\partial_i \left( x^j \,B^{k 0 i} (x) \right) \,|\, 0 \rangle
 \nonumber \\
 &=& \left\{ \begin{array}{l}
 \! \hspace{6mm} 0 \hspace{47mm} \mbox{for} \ p = 0, \\
 \! \lim_{R \rightarrow \infty} \,\int \,d^4 x \,\alpha (x_0) \,f_R (\bm{x}) \,e^{\,i \,p_\mu \,x^\mu} 
 \\
 \hspace{17mm} \times \ i \,p_i \,\langle p \,|\, x^j \,B^{k 0 j} (0) \,|\,0 \rangle \hspace{2mm}
 \mbox{for} \ p \neq 0. \\
 \end{array} \right. \ \ \ \ \ 
\end{eqnarray}

This answer clearly contradicts the formula (\ref{Eq:Basic_Lowdon}) given by Lowdon. 
Note that, if our formula is correct, it immediately follows that the nucleon forward matrix
element of any surface integral term always vanishes. In particular, we have  
\begin{eqnarray}
 \langle p, s \,|\, S^i_1 \,|\,p, s \rangle \ = \ 0, \\
 \langle p, s \,|\, S^i_2 \,|\,p, s \rangle \ = \ 0 .
\end{eqnarray}
in contradiction with Lowdon's conclusion. It seems that the origin of this discrepancy
can be traced back to the fact that the non-commuting nature of the coordinate
and momentum operators indicated by Eq.(\ref{Eq:C.R.}) was overlooked in Lowdon's analysis.  
In fact, if we had discarded it, the $S_1$ term would become
\begin{equation}
 S_1 \ = \ \int \,d^3 x \,e^{\,i \,p_\mu \,x^\mu} \,i \,p_i \,
 \langle p \,|\, x^j \,B^{k 0 i} (0) \,|\, 0 \rangle .
\end{equation}
Combining this result with that for the $S_3$ term, one would then be led to
Lowdon's basic relation (\ref{Eq:Basic_Lowdon}).  We therefore conclude that,
although based on somewhat delicate plane-wave formalism, the statement in
the original analysis by Jaffe-Manohar is basically correct \cite{JM1990}.
This means that
the nucleon forward matrix elements of the surface integral terms 
vanish identically, thereby enabling us to obtain a physically meaningful 
nucleon spin sum rule.

\section{The role of gluon topology on the boundary terms}
\label{sec:3}

Despite the early establishment of the concept of the gluon distribution functions
in the field of deep-inelastic-scatter-\\ing physics,
how we can define the longitudinally polarized gluon distribution function
$\Delta g (x)$ in a gauge-invariant manner remains to be a fairly delicate issue.
In fact, the difficulty had first appeared in the {\it absence} of the twist-2
and gauge-invariant local gluon operator corresponding to the 1st moment of 
$\Delta g (x)$ in the operator-product-expansion (OPE)
framework \cite{Sasaki1975},\cite{AR1975}. 
To get a clear answer to our problem, i.e. the role of surface terms in the
definition of $\Delta g (x)$, we need to make clear all the delicacies left
in the past attempts to provide a satisfactory definition of it.

Following the logical steps for providing a reasonable definition of the unpolarized 
gluon distribution in the pioneering work by Collins and Soper \cite{CS1982}, 
Bashinsky and Jaffe start with the following plausible field-theoretical definition 
of $\Delta g (x)$ in the light-cone (LC) gauge \cite{BJ1998} : 
\begin{eqnarray}
 \Delta g (x) &=& \frac{1}{2} \,\int \,\frac{d \xi^-}{2 \,\pi} \,
 e^{\,i \,x \,P^+ \,\xi^-} \nonumber \\
 &\times& \langle P \,|\, 2 \,\mbox{Tr} \,
 \left\{ F^{+ \lambda} (0) \,\epsilon^{+-}{}_\lambda{}^\nu
 \,A_\nu (\xi^-) \right\} |\,P \rangle . \ \ \ \ 
\end{eqnarray}
Here, the indices $\pm$ in the antisymmetric Levi-Civita symbol stand for the
two light-like components, defined as $V^\pm = (V^0 \pm V^3) / \sqrt{2}$ for
any four-vector $V^\mu$. On the other hand, 
the repeated indices $\lambda$ and $\nu$ represent the standard Lorentz
indices, which are practically restricted to take two transverse components,
i.e. $\lambda, \,\nu = 1 \ \mbox{or} \ 2$.

Substituting $e^{\,i \,x \,P^+ \,\xi^-}$ by $( i \,x \,P^+)^{-1} \,(\partial / \partial \xi^-) \,
e^{\,i \,x \,P^+ \,\xi^-}$, integrating by parts, and using the identity 
$\frac{\partial}{\partial \xi^-} \,A_\nu = F^+{}_\nu = - F_\nu{}^+$ that
holds in the LC gauge, they rewrite the above expression in the following form : 
\begin{eqnarray}
 \Delta g (x) &=& \frac{i}{2 \,P^+ \,x} \,\int \,\frac{d \xi^-}{2 \,\pi} \,
 e^{\,i \,x \,P^+ \,\xi^-} \nonumber \\
 &\,& \hspace{-15mm} \times \, 
 \langle P \,|\, 2 \,\mbox{Tr} \left\{ F^{+ \lambda} (0) \,
 \tilde{F}_\lambda{}^+ (\xi^-) \right\} | \,P \rangle 
 +  
 \left[ \left. \left( \cdots \right) 
 \right|^{\xi^- = + \,\infty}_{\xi^- = - \,\infty} \right]. \hspace{6mm} \label{Eq:Surface_BJ}
\end{eqnarray}
In the above expression, $\tilde{F}^{\mu \nu \alpha \beta} \equiv \frac{1}{2} \,
\epsilon^{\mu \nu \alpha \beta} \,F_{\alpha \beta}$ is 
the dual field-strength tensor, while the second term represents the surface terms 
resulting from the partial integration.
This expression is still incomplete in several respects. First, although Bashinsky-Jaffe
gave a plausible argument that the surface terms are likely to vanish, stronger support
is necessary especially in view of all the subtleties we have already pointed out. 
Second, although the main term of (\ref{Eq:Surface_BJ}) is expressed only with the gluon
field-strength tensor, it still is not invariant under general gauge transformations. 
The reason is the space-time nonlocality of the relevant gluon correlator as well as the
following gauge transformation properties of field-strength tensors :  
\begin{eqnarray}
 F^{+ \lambda} (0) \ &\rightarrow& \ U (0) \,F^{+ \lambda} (0) \,U^\dagger (0), \\
 \tilde{F}_\lambda^+ (\xi^-) \ &\rightarrow& \ U (\xi^-) \,\tilde{F}_\lambda^+ (\xi^-) \,
 U^\dagger (\xi^-).
\end{eqnarray}
Third, the above definition of $\Delta g (x)$ has a $1 / x$ singularity at $x = 0$.
The principal-value prescription, which means the replacement 
$1 / x \rightarrow P \,(1 / x)$, is frequently adopted \cite{KT1999}, 
but its physics foundation is not 
necessarily clear enough.
As we shall see below, all these problems are intricately interrelated.
This means that, to get a completely satisfactory gauge-invariant definition of 
the longitudinally polarized gluon distribution, we are obliged to resolve all these 
delicacies simultaneously as well as without ambiguity.  

To answer the question posed above, we find it very enlightening to remember the 
argument by Belistky, Ji, and Yuan on the transverse-momentum-dependent 
(TMD) \\
quark distribution functions \cite{BJY2003}.
(See also \cite{BMP2003}.)
As they stressed, the TMD quark distributions are known to be generally
{\it process-dependent} quantities. A theoretically satisfactory  definition of the 
TMD quark distribution corresponding to semi-inclusive DIS processes was shown
to be given as
\begin{eqnarray}
 q_{DIS} (x, \bm{k}_\perp) &=& \frac{1}{2 \,P^+} \,\int \,\frac{d \xi^-}{2 \,\pi} \,
 \int \,\frac{d^2 \bm{\xi}_\perp}{(2 \,\pi)^2} \,
 e^{\,i \,x \,P^+ \,\xi^- - i \,\bm{k}_\perp \cdot \bm{\xi}_\perp} \nonumber \\
 &\,& \hspace{-22mm} \times \, \langle P \,|\, \bar{\psi} (0^-, \bm{0}_\perp) \,
 {\cal W}_C [0^-, \bm{0}_\perp \,;\,\xi^-, \bm{\xi}_\perp] \,\psi (\xi^-, \bm{\xi}_\perp) \,
 |\,P \rangle , \ \ \ \ 
\end{eqnarray}
where
\begin{eqnarray}
 {\cal W}_C [0^-, \bm{0}_\perp \,;\, \xi^-, \bm{\xi}_\perp]
 &\equiv& {\cal L} [0^-, \bm{0}_\perp \,;\, + \,\infty^-, \bm{0}_\perp] \,
 \nonumber \\
 &\,& \hspace{-34mm} \times \ 
 {\cal L} [+ \,\infty^- , \,\bm{0}_\perp \,;\, + \,\infty^-, \bm{\xi}_\perp] \,
 {\cal L} [+ \,\infty^-, \bm{\xi}_\perp \,;\, \xi^-, \bm{\xi}_\perp] , \ \ \ 
\end{eqnarray}
is the Wilson line, also called the gauge link, representing the 
future-pointing staple-like 
LC path as illustrated in Fig.1(a). 
(In the present paper, we reserve the notation ${\cal L} [\eta^-, \bm{\eta}_\perp \,;\,
\xi^-, \bm{\xi}_\perp]$ as representing a Wilson-line that connects
the two space-time points $(\xi^-, \bm{\xi}_\perp)$ and
$(\eta^-, \bm{\eta}_\perp)$ with a straight-line path.)

\vspace{4mm}

\begin{figure*}[ht]
\begin{center}
 \includegraphics[width=0.6\linewidth]{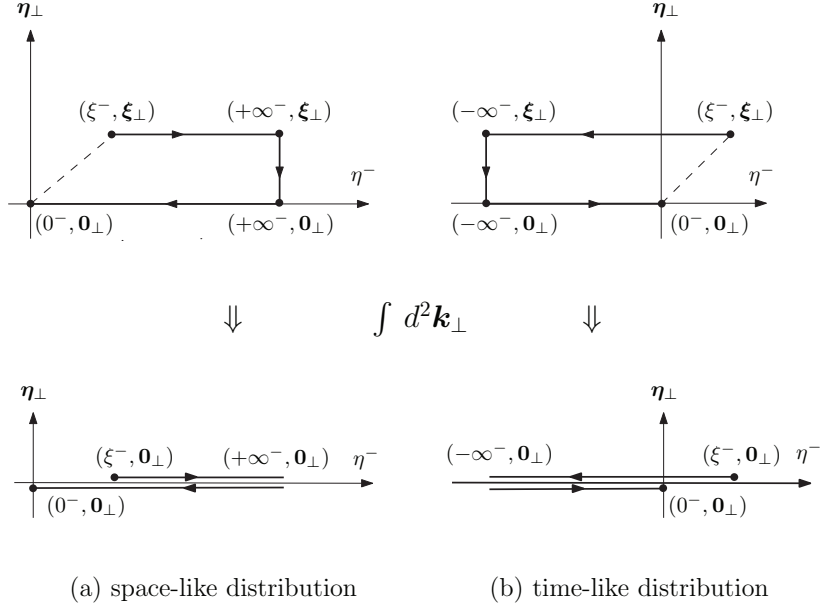}
\caption{The left and right figures in the upper panel respectively stand for
the future-pointing staple-like LC path and the past-pointing staple-like LC
path simulating the final-state interaction in semi-inclusive DIS processes and
the intial-state interaction in Drell-Yan processes. The lower panel represents
the gauge-link structures obtained after integrating out the transverse momentum $\bm{k}_\perp$.} 
\label{Fig:fig1}
\end{center}
\end{figure*}

\vspace{2mm}

On the other hand, the TMD quark distribution 
corresponding to Drell-Yan processes is defined by
\begin{eqnarray}
 q_{DY} (x, \bm{k}_\perp) &=& \frac{1}{2 \,P^+} \,\int \frac{d \xi^-}{2 \,\pi}
 \int \frac{d^2 \bm{\xi}_\perp}{(2 \,\pi)^2} \,
 e^{\,i \,x \,P^+ \,\xi^- - i \,\bm{k}_\perp \cdot \bm{\xi}_\perp} \nonumber \\
 &\,& \hspace{-22mm} \times \, \langle P \,|\, \bar{\psi} (0^-, \bm{0}_\perp) \,
 {\cal W}_{C^\prime} [0^-, \bm{0}_\perp \,;\,\xi^-, \bm{\xi}_\perp] \,\psi (\xi^-, \bm{\xi}_\perp) \,
 |\,P \rangle , \ \ \ \ \ 
\end{eqnarray}
where
\begin{eqnarray}
 {\cal W}_{C^\prime} [0^-, \bm{0}_\perp \,;\, \xi^-, \bm{\xi}_\perp]
 &\equiv& {\cal L} [0^-, \bm{0}_\perp \,;\, - \,\infty^-, \bm{0}_\perp] \,
 \nonumber \ \ \ \\
 &\,& \hspace{-32mm} \times \ 
 {\cal L} [- \,\infty^-, \,\bm{0}_\perp \,;\, - \,\infty^-, \bm{\xi}_\perp] \,
 {\cal L} [- \,\infty^-, \bm{\xi}_\perp \,;\, \xi^-, \bm{\xi}_\perp] , \ \ \ \ \ 
\end{eqnarray}
is the Wilson line representing the past-pointing staple-like LC path as
illustrated in Fig.1(b).

When integrated over the transverse momentum, one gets the following expressions
for the quark distribution for DIS processes
\begin{eqnarray}
 q_{DIS} (x) &=& \frac{1}{2\,P^+} \,\int \,\frac{d \xi^-}{2 \,\pi} \,e^{\,i \,x \,P^+ \,\xi^-} \,
 \langle P \,|\,\bar{\psi} (0^-, \bm{0}_\perp) \,\gamma^+ \, \nonumber \ \ \ \\
 &\,& \hspace{-10mm} \times \ {\cal L} [0^-, \bm{0}_\perp \,;\, + \,\infty^-, \bm{0}_\perp] \,
 {\cal L} [+ \,\infty^-, \bm{0}_\perp \,;\, \xi^-, \bm{0}_\perp] \nonumber \\
 &\,& \hspace{35mm} \times \ \psi (\xi^-, \bm{0}_\perp) \,|\, P \rangle , \ \ \ \ \ 
\end{eqnarray}
and for the distribution for Drell-Yan processes
\begin{eqnarray}
 q_{DY} (x) &=& \frac{1}{2\,P^+} \,\int \,\frac{d \xi^-}{2 \,\pi} \,e^{\,i \,x \,P^+ \,\xi^-} \,
 \langle P \,|\,\bar{\psi} (0^-, \bm{0}_\perp) \,\gamma^+ \, \nonumber \ \ \ \ \ \\
 &\,& \hspace{-10mm} \times \ {\cal L} [0^-, \bm{0}_\perp \,;\, - \,\infty^-, \bm{0}_\perp] \,
 {\cal L} [- \,\infty^-, \bm{0}_\perp \,;\, \xi^-, \bm{0}_\perp] \nonumber \\
 &\,& \hspace{35mm} \times \ \psi (\xi^-, \bm{0}_\perp) \,|\, P \rangle . \ \ \ 
\end{eqnarray}
As Belistky, Ji, and Yuan pointed out, the unitarity of the gauge link implies that
\begin{eqnarray}
 &\,& {\cal L} [0^-, \bm{0}_\perp \,;\,+ \,\infty^-, \bm{0}_\perp] \,
 {\cal L} [+ \,\infty^-, \bm{0}_\perp \,;\,\xi^-, \bm{0}_\perp] \ \ \ \ \nonumber \\
 &=&  {\cal L} [0^-, \bm{0}_\perp \,;\,- \,\infty^-, \bm{0}_\perp] \,
 {\cal L} [- \,\infty^-, \bm{0}_\perp \,;\,\xi^-, \bm{0}_\perp] \nonumber \\
 &=& {\cal L} [0^-, \bm{0}_\perp \,;\,\xi^-, \bm{0}_\perp] , \ \ \ \ \ 
\end{eqnarray}
so that one eventually obtains
\begin{eqnarray}
 q_{DIS} (x) &=& \ q_{DY} (x) \ = \ 
 \frac{1}{2 \,P^+} \,\int \,\frac{d \xi^-}{2 \,\pi} \,e^{\,i \,x \,P^+ \,\xi^-} 
 \nonumber \\
 &\,& \hspace{-18mm} \times \, \langle P \,|\, \bar{\psi} (0^-, \bm{0}_\perp) \,\gamma^+ \,
 {\cal L} [0^-, \bm{0}_\perp \,;\, \xi^-, \bm{0}_\perp] \,
 \psi (\xi^-, \bm{0}_\perp) \,| \,P \rangle , \ \ \ \ \ 
\end{eqnarray}
which means that both distributions are exactly the same, thereby ensuring the 
universality of the $\bm{k}_\perp$-integrated \\
quark distributions.

One might expect that the same argument can be used also for showing
the universality of the longitudinally polarized gluon distribution $\Delta g (x)$.
Things are not so straightforward because of the $1 / x$ singularity appearing in
the candidates of the gauge-invariant definition of $\Delta g (x)$.
We first recall that the popular choice is to use the principal-value prescription
as advocated by Manohar \cite{Manohar1990},\cite{Manohar1991}, which gives
\begin{eqnarray}
 \Delta g (x) &\equiv& \frac{1}{4 \,P^+} \,\int \,\frac{d \xi^-}{2 \,\pi} \,
 P \,\frac{i}{x} \,e^{\,i \,x \,P^+ \,\xi^-} \nonumber \\
 &\,& \hspace{-8mm} 
 \times \ \langle P S_\| \,|\, 2 \,\mbox{Tr} \,\left\{ \tilde{F}^+_i (0) \,
 {\cal L} [0^-, \xi^-] \,F^{+ i} (\xi^-) \right\} |\, P S_\| \rangle \ \ \ \ \ 
 \nonumber \\
 &+& \  (x \ \rightarrow - \,x) , \ \ \ \ \ 
\end{eqnarray}
where the second term is included to ensure the crossing symmetry of 
$\Delta g (x)$. This is not the only choice, however. 
Possible alternatives would be given by
\begin{eqnarray}
 \Delta g_{post/prior} (x) &\equiv& \frac{1}{4 \,P^+} \,\int \,\frac{d \xi^-}{2 \,\pi} \,
 \frac{i}{x \mp i \,\epsilon} \,e^{\,i \,x \,P^+ \,\xi^-} \nonumber \\
 &\,& \hspace{-24mm} 
 \times \ \langle P S_\| \,|\, 2 \,\mbox{Tr} \,\left\{ \tilde{F}^+_i (0^-) \,
 {\cal L} [0^-, \xi^-] \,F^{+ i} (\xi^-) \right\} \,|\, P S_\| \rangle \ \ \ \ \ 
 \nonumber \\
 &\,& \hspace{-18mm} + \  (x \ \rightarrow - \,x) . \ \ \ \ \ 
\end{eqnarray}
Here the post and the prior forms of $\Delta g (x)$ respectively correspond
to the choices $1/x \rightarrow 1 / (x - i \,\epsilon)$ and 
$1/x \rightarrow 1 / (x + i \,\epsilon)$.
The reason of this naming is as follows. Using the mathematical identities
\begin{equation}
 \int_{- \,\infty}^\infty \,d x \,\frac{i}{x - i \,\epsilon} \,
 e^{\,\pm \,i \,x \,P^+ \,\xi^-} \ = \ 
 - \,2 \,\pi \,\theta (\pm \,\xi^-) ,
\end{equation}
and
\begin{equation}
 \int_{- \,\infty}^\infty \,d x \,\frac{i}{x + i \,\epsilon} \,
 e^{\,\pm \,i \,x \,P^+ \,\xi^-} \ = \ + \,2 \,\pi \,\theta (\mp \,\xi^-) ,
\end{equation}
with $\theta (x)$ being the standard step function, the first moments of
$\Delta g_{post} (x)$ and $\Delta g_{prior} (x)$ respectively reduce to
\begin{eqnarray}
 \Delta G_{post}^{(1)} &\equiv& \int_{- \,\infty}^\infty d x \,\Delta g_{post} (x) 
 \, = \, \frac{1}{2 \,P^+} \int d \xi^- \left( - \,\theta (\xi^-) \right) \nonumber \\
 &\,& \hspace{-13mm} \times \, \langle P S_\| \,|\, 2 \,\mbox{Tr} \,\left\{ \tilde{F}^+_i (0^-) \,
 {\cal L} [0^- , \xi^-] \,F^{+ i} (\xi^-) \right\} |\, P S_\| \rangle , \ \ \ \ \label{Eq:Gpost} 
\end{eqnarray}
and
\begin{eqnarray}
 \Delta G_{prior}^{(1)} &\equiv& 
 \int_{- \,\infty}^\infty d x \,\Delta g_{prior} (x) \, = \, 
 \frac{1}{2 \,P^+} \int d \xi^- \left( \theta (- \,\xi^-) \right) \nonumber \\
 &\,& \hspace{-15mm} \times \, \langle P S_\| \,|\, 2 \,\mbox{Tr} \,\left\{ \tilde{F}^+_i (0^-) \,
 {\cal L} [0^- , \xi^-] \,F^{+ i} (\xi^-) \right\} |\, P S_\| \rangle. \label{Eq:Gprior} 
\end{eqnarray}
The appearance of the factors $\theta (\xi^-)$ in (\ref{Eq:Gpost}) and 
$\theta (- \,\xi^-)$ in (\ref{Eq:Gprior}) is thought
to be a reminiscence of the future-pointing and past-pointing gauge-link structures
illustrated in \\
Fig.\ref{Fig:fig1}(a) and Fig.\ref{Fig:fig1}(b), 
which are interpreted to simulate the final-state
interaction in the DIS processes and the initial-state interaction in the
Drell-Yan processes. This implies the identification
\begin{eqnarray}
 \Delta g_{post} (x) \ &\longleftrightarrow& \ \Delta g_{DIS} (x), \\
 \Delta g_{prior} (x) \ &\longleftrightarrow& \ \Delta g_{DY} (x).
\end{eqnarray}
The standard principal-value prescription just amounts to taking an average of 
the post form and the prior form : 
\begin{equation}
 \Delta g_{PV} (x) \ = \ \frac{1}{2} \,\left( \Delta g_{post} (x)
 \ + \ \Delta g_{prior} (x) \right) .
\end{equation}
We emphasize that the equality of the above
three distributions is not self-evident from the beginning.
Thus, the logical consequence of the consideration above is that, as candidates of 
physically meaningful definitions of longitudinally
polarized gluon distribution, it is legitimate to start with either of the post form
or the prior form depending on the process that we are considering, as Belitsky, Ji
and Yuan did in their proof of universality of the $\bm{k}_\perp$-integrated 
quark distribution function \cite{BJY2003}.

Let us now investigate these two physics-based distributions in some detail.
With use of the familiar mathematical identity
\begin{equation}
 \frac{1}{x \mp i \,\epsilon} \ = \ P \,\frac{1}{x} \ \pm \ i \,\pi \,\delta (x) ,
\end{equation}
and
\begin{equation}
 \int_{- \,\infty}^{+ \,\infty} \,d x \,\,P \,\frac{i}{x} \,\,
 e^{\,\pm \,i \,x \,P^+ \,\xi^-} \ = \ 
 \mp \,\pi \,\epsilon (\xi^-) ,
\end{equation}
we find that the post and prior form of $\Delta g (x)$ can be expressed as
\begin{equation}
 \Delta g_{post/prior} (x) \ = \ \Delta g_{PV} (x) \ \mp \ c \,\delta (x),
\end{equation}
where
\begin{eqnarray}
 \Delta g_{PV} (x) &=& \frac{1}{2 \,P^+} \,\int_{- \,\infty}^\infty \,
 \frac{d \xi^-}{2 \,\pi} \,P \,\frac{i}{x} \,e^{\,i \,x \,P^+ \,\xi^-} \nonumber \\
 &\,& \hspace{-16mm} \times \, 
 \langle P S_\| \,|\,2 \,\mbox{Tr} \left\{ \tilde{F}^+_i (0^-) \,{\cal L} [0^-, \xi^-] \,
 F^{+i} (\xi^-) \,\right\} |\, P S_\| \rangle , \ \ \ \ \ 
\end{eqnarray}
and
\begin{eqnarray}
 c \ &=& \ \frac{1}{4 \,P^+} \,\int_{- \,\infty}^\infty \,d \xi^- \,
 \langle P S_\| \,|\,2 \,\mbox{Tr} \left\{ \tilde{F}^+_i (0^-) \,{\cal L} [0^-, \xi^-] 
 \right. \ \ \ \nonumber \\
 &\,& \hspace{38mm} \times \ \left. F^{+i} (\xi^-) \,\right\} \,|\, P S_\| \rangle .
\end{eqnarray}
A key question now is therefore whether the constant $c$ vanishes or not at all.

Before answering this question, we think it instructive to look into several
lower moments of $\Delta g_{post/prior} (x)$. For the first, the second, and the
third moments, we find that
\begin{eqnarray}
 \Delta G^{(1)}_{post/prior} &\equiv& \int \,d x \,\,\Delta g_{post/prior} (x) \nonumber \\
 &\,& \hspace{-20mm} \ = \ 
 \frac{1}{2 \,P^+} \,\int d \xi^- \,
 \left( - \,\frac{1}{2} \,\epsilon (\xi) \right) \nonumber \\
 &\,& \hspace{-18mm} \times \ \langle P S_\| \,|\, 2 \,\mbox{Tr} \left\{ \tilde{F}^+_i (0^-) \,
 {\cal L} [0^-, \xi^-] \,F^{+ i} (\xi^-) \right\} \,|\,P S_\| \rangle \nonumber \\
 &\,&  \hspace{-18mm} \mp \ c , \ \ \ \ \ \\
 \Delta G^{(2)}_{post/prior} &\equiv& \int \,d x \,\,x \,\Delta g_{post/prior} (x) \nonumber \\
 &\,& \hspace{-18mm} = \ \frac{i}{4 \,(P^+)^2} \,\,
 \langle P S_\| \,|\, 2 \,\mbox{Tr} \left\{ \tilde{F}^+_i (0^-) \,F^{+ i} (0^-) \right\} \,|\,P S_\| \rangle
 \nonumber \\
 &\,& \hspace{-18mm} - \ \frac{i}{4 \,(P^+)^2} \,\,
 \langle P S_\| \,|\, 2 \,\mbox{Tr} \left\{ \tilde{F}^+_i (0^-) \,F^{+ i} (0^-) \right\} \,|\,P S_\| \rangle
 \nonumber \\
 &\,& \hspace{-19mm}
 \ = \ 0 , \\
 \Delta G^{(3)}_{post/prior} &\equiv& \int \,d x \,\,x^2 \,\Delta g_{post/prior} (x) \nonumber \\
 &\,& \hspace{-26mm} = - \,\frac{1}{2 \,(P^+)^3} \langle P S_\| |\, 2 \,\mbox{Tr} \left\{
 \tilde{F}^+_i (0^-) D^+ F^{+ i} (0^-) \right\} | P S_\| \rangle, \ \ \ \ 
\end{eqnarray}
where $D^+ = \partial^+ - i \,g \,A^+$ is the $+$-component of the covariant derivative.

We first point out that the second moment vanishes as a cancellation of two terms
emerging from the crossing symmetry of the DIS amplitude, each term of which
is expressed as a nucleon matrix element of the topological charge operator
of the gluon. The third moment just coincides with the expression anticipated
from the operator-product expansion \cite{Sasaki1975},\cite{AR1975}. 
This is naturally the case also for
higher moments. As anticipated, a subtlety remains only for the
first moment, which is related to the net gluon spin contribution to the
nucleon spin sum rule.

To pursue further the subtlety of the first moment of the longitudinally polarized gluon
distribution, it is useful to inspect what form
it reduces to in a special gauge in our problem, i.e. in the LC gauge.
Using the relations ${\cal L} [0^-, \xi^-] = 1$
and $F^{+i} (\xi^-) = \frac{\partial}{\partial \xi^-} \,A^i (\xi^-)$ that holds in the
LC gauge, we obtain
\begin{eqnarray}
 \Delta G^{(1)}_{post/prior} \ = \ \Delta G^{(1)}_{PV} \ \mp \ c ,
\end{eqnarray}
where
\begin{eqnarray}
 \Delta G^{(1)}_{PV} &=& \frac{1}{2 \,P^+} \,
 \langle P S_\| \,|\, 2 \,\mbox{Tr} \left\{ \tilde{F}^+_i (0) \right. \nonumber \\
 &\,& \hspace{-15mm} \times \left. \left[ A^i (0^-)  -  \frac{1}{2} \,
 \left( A^i (+ \,\infty^-)  +  A^i (- \,\infty^-) \right) \right] \right\}
 | P S_\| \rangle , \ \ \ \ \ 
\end{eqnarray}
and
\begin{eqnarray}
 c \ &=& \ \frac{1}{2 \,P^+} \,
 \langle P S_\| \,|\, 2 \,\mbox{Tr} \left\{ \tilde{F}^+_i (0) \right. \nonumber \\
 &\,& \hspace{8mm} \times \,\left. \frac{1}{2} \,
 \left( A^i (+ \,\infty^-) - A^i (- \,\infty^-) \right) \right\} 
 |\,P S_\| \rangle . \ \ \ \ 
\end{eqnarray}
Combining these, we therefore get
\begin{eqnarray}
 \Delta G^{(1)}_{post} &=& \frac{1}{2 \,P^+} \,
 \langle P S_\| \,|\, 2 \,\mbox{Tr} \left\{ \tilde{F}^+_i (0) \right. \nonumber \\
 &\,& \hspace{3mm} \times \,\left.
 \left( A^i (0^-) - A^i (+ \,\infty^-) \right) \right\} 
 |\,P S_\| \rangle , \ \ \ \\
 \Delta G^{(1)}_{prior} &=& \frac{1}{2 \,P^+} \,
 \langle P S_\| \,|\, 2 \,\mbox{Tr} \left\{ \tilde{F}^+_i (0) \right. \nonumber \\
 &\,& \hspace{3mm} \times \,\left.
 \left( A^i (0^-) - A^i (- \,\infty^-) \right) \right\} 
 |\,P S_\| \rangle , \ \ \ 
\end{eqnarray}
We recall that the first moment in the post form essentially
coincides with the expression written down by Bass inspired
by the paper of Manohar \cite{Manohar1990}.
 (See Eq.(90) of Ref.\cite{Bass2005}.)
Manohar conjectured that the gluon correlation function
$\langle P S_\| \,|\, \mbox{Tr} \left\{ \tilde{F}^+_i (0^-) \,A^i (\xi^-) \right\}
\,|\,P S_\| \,\rangle$ would vanish as $\xi^- \rightarrow \pm \,\infty$.
On the other hand, Bass suspects that, owing to the nontrivial topology
of the gluon configuration in the QCD vacuum, this surface term might
not necessarily vanish but it rather generates
a delta-function type singularity at $x = 0$ in the longitudinally
polarized gluon distribution function $\Delta g(x)$. 
It must be stressed that, if $\Delta g(x)$
has such a singularity, it would invalidate a naive partonic
sum rule for the net nucleon spin. It is therefore of vital importance to
carefully check whether this surface term contribution vanishes or not.

Widely known remarkable fact in the LC gauge is that, 
$A^i (+ \,\infty^-)$ and $A^i (- \,\infty^-)$
cannot be set to zero simultaneously, because of the Gauss law
constraint $A^i (+\,\infty^-) - A^i (- \,\infty^-) = 
\int_{- \,\infty^-}^{+ \,\infty^-} \,d \eta^- \,F^{+ i} (\eta^-) \neq 0$.
Frequently used boundary condition (b.c.) for the gluon field is either of the 
following three : 
\begin{eqnarray}
 &\,& A^i (+ \,\infty^-) \, = \, 0 \hspace{19mm} \ : \ \mbox{advanced b.c.} ,\\
 &\,& A^i (- \,\infty^-) \, = \, 0 \hspace{19mm} \ : \ \mbox{retarded b.c.} ,\\
 &\,& A^i (+ \,\infty^-)  +  A^i (- \,\infty^-)  =  0 \ : \  
 \mbox{antisymmetric b.c.} \ \ \ \ \ \ 
\end{eqnarray}

According to Hatta \cite{Hatta2011}, how to treat the $1 / x$ singularity 
in the gluon distribution
is connected with the choices of the boundary condition for the gluon field.
He defines the so-called physical component of the gluon field which transforms
covariantly under gauge transformation, for three choices of the boundary
condition. For retarded and advanced boundary conditions, it is defined as
\begin{eqnarray}
 A^{(Ret/Adv)\, i}_{phys, a} (0) &=& \int \,d x \,\int_{- \,\infty}^{+ \,\infty} \,
 \frac{d \eta^-}{2 \,\pi} \,\frac{i}{x \mp i \,\epsilon} \,e^{\,i \,x \,P^+ \,\eta^-} 
 \nonumber \\
 &\,& \hspace{15mm} \times \ 
 {\cal L}_{a b} [0^-, \eta^-] \,F^{+ i}_b (\eta^-) \nonumber \\
 &=& \int_{- \,\infty}^{+ \,\infty} \,d \eta^- \,
 \left( \mp \,\theta (\pm \,\eta^-) \right) \
 \nonumber \\
 &\,& \hspace{15mm} \times \ 
 {\cal L}_{a b} [0^-, \eta^-] \, F^{+ i}_b (\eta^-) ,
\end{eqnarray}
while, for the antisymmetric (AS) boundary condition, it is defined as
\begin{eqnarray}
 A^{(AS) \,i}_{phys, a} (0) &=& \int \,d x \,\int_{- \,\infty}^{+ \,\infty} \,
 \frac{d \eta^-}{2 \,\pi} \,P \,\frac{i}{x} \,e^{\,i \,x \,P^+ \,\eta^-} \nonumber \\
 &\,& \hspace{18mm} \times \,
 {\cal L}_{a b} [0^-, \eta^-] \,F^{+ i}_b (\eta^-) \nonumber \\
 &=& \int_{- \,\infty}^{+ \,\infty} \,d \eta^- \,
 \left( - \frac{1}{2} \,\epsilon (\eta^-) \right) \nonumber \\
 &\,& \hspace{18mm} \times \,{\cal L}_{a b} [0^-, \eta^-] \,
 F^{+ i}_b (\eta^-) . \ \ \ \ \ \ 
\end{eqnarray}
For each definition of the physical component, the gluon spin, or more
precisely the first moment of the longitudinally polarized gluon distribution,
takes the following form :
\begin{equation}
 \Delta G^{(1)} = \frac{1}{2 \,P^+} \,\langle P S_\parallel \,|\,
 2 \,\mbox{Tr} \left\{ \tilde{F}^+_i (0) \,A^i_{phys} (0) \right\}
 | P S_\parallel \rangle ,
\end{equation}
where $A^i_{phys}$ is the physical component in any of the retarded,  
advanced, or antisymmetric boundary conditions.

%

As is discussed above, however, the choice of the post or prior form in the 
definition of $\Delta g (x)$ is a physics-based operation related to
the space-time structure of the Wilson line, which simulates the
final-state interaction in the DIS processes or the initial-state interaction
in the Drell-Yan processes. This choice can in principle be independent of 
the choice of the boundary condition for the gluon field at the LC spatial
infinity. In fact, in either choice of the post or prior form, we have a
freedom to work in the antisymmetric boundary condition for the gluon
field. In fact, this is what Burkardt did 
in his study of the role of final-state interaction in parton orbital 
angular momentum \cite{Burkardt2013}. 
(In \cite{Waka2015}, we tried to prove the nonexistence of the
delta-function singularity in $\Delta g(x)$. However, the proof given
there is incomplete, because there was a confusion between
the choices of the boundary condition for the gluon field and the post-
and prior-form definitions of $\Delta g (x)$.)

From a practical standpoint, i.e. if one intends to solve the 
bound-state problem of the nucleon as a coupled quark-gluon system 
in some way, the antisymmetric boundary condition
would be the most natural and convenient choice.
For example, within the framework of the light-front quantization, 
Zhang and Harindranath advocated to use the antisymmetric boundary condition 
as a natural choice to fix the residual gauge freedom \cite{ZH1993}. 
They also claim that
the topological winding number of the gluon field is fixed by the non-zero
boundary value $A^i (+ \,\infty^-) = - \,A^i (- \,\infty^-)$.
In fact, they showed that the winding number $\Delta Q_5$ of the gluon field
can be expressed in the following form : 
\begin{eqnarray}
 \Delta Q_5 \ &\equiv& \ n_f \,\frac{g^2}{8 \,\pi^2} \,\int_M \,
 \mbox{Tr} \,\left( \tilde{F}^{\mu \nu} \,F_{\mu \nu} \right) \nonumber \\
 &\,& \hspace{-14mm} = - \,n_f \,\frac{g^2}{\pi^2} \,\int \,d \xi^+ \,d^2 \bm{\xi}_\perp
 \left. \mbox{Tr} \left( A^- \,[\,A^1, A^2] \,\right) \, 
 \right|_{\xi^- = - \,\infty}^{\xi^- = + \,\infty}, \ \ \ \ \label{Eq:topological}
\end{eqnarray}
with $n_f$ being the number of quark flavors.
Note that, since the $A^-$ component in the light-front quantization scheme
is not an independent field, $\left. A^-_a \right|_{- \,\infty}^{+ \,\infty}$ is
after all determined by the surface values $\left. A^{1,2}\right|_{- \,\infty}^{+ \,\infty}$
of the two independent fields. This in turn means that $\Delta Q_5$ in 
Eq.(\ref{Eq:topological}) is determined solely by the surface values 
$\left. A^{1,2}\right|_{- \,\infty}^{+ \,\infty}$ of the two independent fields.

As a matter of course, despite the practical advantage of using the antisymmetric 
boundary condition in solving the bound state problem, the choice of boundary 
condition in the LC gauge is in principle 
arbitrary, and one can choose other two boundary conditions as well.
The only thing one must be careful about is that
the latter choices would bring about some complexity, because
the corresponding bound state wave function of the nucleon generally 
acquires complex phase \cite{BJY2003}. 
Despite this complexity, if everything is treated consistently, 
the final physical prediction for a gauge-invariant quantity is naturally expected to be 
independent of the choice of boundary condition within the LC gauge.

Now we are in a position to answer our central question.
Does the coefficient $c$ of the $\delta (x)$ term vanish or not ?
First, we point out that, by using the translational invariance, 
$c$ can be rewritten in the following form : 
\begin{eqnarray}
 c \ &=& \ \frac{1}{2 \,P^+} \,\langle P S_\| \,|\, 2 \,\mbox{Tr} \left\{
 \tilde{F}^+_i (0^-) \right. \nonumber \\
 &\,& \hspace{10mm} \times \,
 \left. \frac{1}{2} \,\left( A^i (+ \,\infty) \, - \, A^i (- \,\infty) \right)
 \right\} \,|\, P S_\| \rangle \nonumber \\
 &=& \frac{1}{2 \,P^+} \,\langle P S_\| \,|\, 2 \,\mbox{Tr} \left\{
 \tilde{F}^+_i (\xi^-) \right. \nonumber \\
 &\,& \hspace{10mm} \times \,
 \left. \frac{1}{2} \,\left( A^i (+ \,\infty) \, - \, A^i (- \,\infty) \right)
 \right\} \,|\, P S_\| \rangle \nonumber \\
 &=& \left( 1 \,/\,  \left( 2 \,P^+ \,\int \,d \xi^- \right) \right). 
 \nonumber \\
 &\times& \langle P S_\| \,|\, \int \,d \xi^- \,2 \,\mbox{Tr} \left\{
 \tilde{F}^+_i (\xi^-) \right. \nonumber \\
 &\,& \hspace{10mm} \left. \times \, \frac{1}{2} \,
 \left( A^i (+ \,\infty) \, - \, A^i (- \,\infty) \right)
 \right\} \,| P S_\| \rangle . \ \ \ 
\end{eqnarray}
Bashinsky and Jaffe argued that this expression is expected to vanish as an
infinite volume average of the derivative of a bound function \cite{BJ1998}.
More tangible proof would be the following.
Using the relation $\tilde{F}^+_i = g_{i j} \,\epsilon^{j k}_\perp \,F^{+ k}$ and
$F^{+ k} = \frac{\partial}{\partial \xi^-} \,A^k$, the above expression can be
rewritten as
\begin{eqnarray}
 c \ &=& \ \left( 1 \,/\,  \left( 2 \,P^+ \,\int \,d \xi^- \right) \right) 
 \nonumber \\
 &\,& \times \ 
 g_{i j} \,\epsilon^{j k}_\perp
 \langle P S_\| | \int_{- \,\infty}^\infty d \xi^- \! \frac{\partial}{\partial \xi^-} 
 \,2 \,\mbox{Tr} \left\{ A^k (\xi^-) \right. \nonumber \\
 &\,& \hspace{15mm} \times \,\left.
 \frac{1}{2} \left( A^i (+ \,\infty) \!-\! A^i (- \,\infty) \right) \right\} 
 | P S_\| \rangle . \ \ \ \ \ 
\end{eqnarray}
Since this is proportional to a nucleon forward matrix element of a surface
integral term, it must vanish, as we have argued in the previous section.
We therefore conclude that
\begin{equation}
 c \ = \ 0.
\end{equation}
It means that the longitudinally polarized gluon distribution does not
have a delta-function-type singularity at $x = 0$. This therefore ensures the 
existence of physically meaningful partonic sum rule
for the total nucleon spin.

\section{The role of surface terms on the density-level decomposition of nucleon spin}
\label{sec:4}

So far, our central concern has been the effect of surface terms on the
integrated sum rule of the nucleon spin, and we have concluded that surface
terms do not contribute to the integrated sum rule. 
In most cases, a similar statement holds as well also
for the decomposition of the total angular momentum of the photon into
its spin and orbital parts.
As a matter of course, however, if we are interested in the decomposition
of the total angular momentum at the density level, an unguarded neglect of
the surface terms would not be justified.

In recent papers \cite{Leader2016},\cite{Leader2018}, 
Leader carried out a comparable analysis of two different 
decompositions of the total angular momentum of a free photon beam into 
orbital and spin parts at the density level.
The one is the angular momentum density, which he calls the Poynting
(or Belinfante) version
\begin{equation}
 \bm{j}_{poyn} (\bm{x}) \ = \ \bm{x} \times (\bm{E} \times \bm{B}),
\end{equation}
with the corresponding integrated angular momentum
\begin{equation}
 \bm{J}_{poyn} \ = \ \int \,d^3 x \,\,\bm{x} \times (\bm{E} \times \bm{B}).
\end{equation}
(Here and hereafter, the dielectric constant of vacuum $\epsilon_0$ is 
set to be unity, for simplicity.)
The other is the so-called gauge-invariant-canonical (g.i.c.) version given as
\begin{equation}
 \bm{J}_{g.i.c.} \ = \ \int \,d^3 x \,\,\bm{j}_{g.i.c.} (\bm{x}) ,
\end{equation}
where
\begin{equation}
 \bm{j}_{g.i.c.} (\bm{x}) \ = \ \bm{l}_{g.i.c.} (\bm{x}) \ + \ 
 \bm{s}_{g.i.c.} (\bm{x}) ,
\end{equation}
with
\begin{eqnarray}
 \bm{l}_{g.i.c.} (\bm{x}) &=& E^i (\bm{x}) \,(\bm{x} \times \nabla) \,
 A^i_\perp (\bm{x}), \\
 \bm{s}_{g.i.c.} (\bm{x}) &=& 
 \bm{E} (\bm{x}) \times \bm{A}_\perp (\bm{x}) .
\end{eqnarray}
Here, $\bm{A}_\perp$ represents the transverse component of the vector 
potential $\bm{A}$.
It is a widely-known fact that $\bm{J}_{poyn}$ and $\bm{J}_{g.i.c.}$ generally
differ by a surface integral term (S.T.) as
\begin{equation}
 \bm{J}_{poyn} \ = \ \bm{J}_{g.i.c.} \ + \ \mbox{S. T.} \label{Eq:Surface_Leader}
\end{equation}
As we have already pointed out, the surface integral term vanishes in
most circumstances, in which the photon fields vanish at the spatial infinity, 
which therefore dictates that $\bm{J}_{poyn} \ = \ \bm{J}_{g.i.c.}$.
(One should keep in mind the fact, however,  that there are some
unusual cases where the 
above condition is not satisfied. For example, it was discussed in \cite{OA2014}
that the surface term vanishes for ``bullet-like'' photon beam, but it does not
for ``pencil-like'' photon beam that has an infinite extent along the direction
of the beam.)
However, it never means that these two angular momenta are the same
at the density level, that is, we would generally have 
$\bm{j}_{poyn} \neq \bm{j}_{g.i.c.}$.
As a concrete example, Leader considered a monochromatic paraxial electric field
propagating in the $z$-direction \cite{ABSW1992}, which is represented as
\begin{equation}
 \bm{E} (\bm{x}) = \left( u (\bm{x}), \,v (\bm{x}), \,\frac{- \,i}{k} 
 \left( \frac{\partial u}{\partial x} + \frac{\partial v}{\partial y} \right) \right) 
 e^{\,i \,(k \,z \,- \,\omega \,t)} , 
\end{equation}
with the choice
\begin{equation}
 v (\bm{x}) \ = \ i \,\sigma \,u (\bm{x}) \ \ \ \mbox{with} \ \ 
 \sigma \ = \ \pm \,1 .
\end{equation}
The above choice of field approximately corresponds to right and left
circular polarization.
For the function $u (\bm{x})$, he chooses the following form : 
\begin{equation}
 u (\rho, \phi, z) \ = \ f (\rho, z) \,e^{\,i \,l \,\phi} ,
\end{equation}
in cylindrical coordinate $(\rho, \phi, z)$ so that it represents a
vortex beam with the azimuthal mode index $l$ (the orbital angular
momentum component along the $z$-direction). For the above photon beam,
Allen et al. had shown that the cycle average of the $z$-component
of the Poynting density, $\langle j_{poyn, z} \rangle$,
per unit power modulo $\epsilon_0 / \omega$, is given by \cite{ABSW1992}
\begin{equation}
 \langle j_{poyn, z} \rangle \ = \ l \,|u|^2 \ - \ \frac{\sigma}{2} \,\rho \,
 \frac{\partial |u|^2}{\partial \rho} .
\end{equation}
On the other hand, Leader showed that the corresponding cycle average
of the g.i.c. version of angular momentum takes the following form \cite{Leader2018} : 
\begin{equation}
 \langle j_{g.i.c., z} \rangle \ = \ l \,|u|^2 \ + \ \sigma \,|u|^2 .
\end{equation}
Comparing these two expressions, he claims that, at the density level,
the g.i.c. version, not the Poynting version, appears to show clean separation 
into the orbital
and spin parts. He also claims that the difference of the spin density
in the two versions can in principle be verified by measuring the
spin angular momentum transfer to the internal angular momentum of
an external atom.

These are reasonable claims as far as the photon laser physics is concerned.
However, in a companion paper \cite{Leader2016}, he made a misleading 
overstatement as if his analysis also resolved a conflict between laser
optics and particle physics. (The latter is obviously concerned with the
nucleon spin decomposition problem.)
The reason why his analysis is never thought to resolve the conflict between 
the laser optics and particle physics is very simple. It is because, while he is 
basically treating a photon beam in free space, a gluon in the nucleon
is not a free particle. To understand the importance of this difference, 
i.e. the difference between free and bound (or interacting) gauge fields,
the nonabelian nature of the gluon is not essential. 
We can stay in the abelian gauge theory, i.e. electrodynamics except that, 
different from Leader, we must consider photons interacting with charged 
particles in a nonperturbative manner.

A crucial point here is that, even in interacting theory, the total
angular momentum of the photon field is given in terms of the Poynting vector
as
\begin{equation}
 \bm{J}^\gamma \ = \ \bm{J}_{poyn} \ =  \ 
 \int \,d^3 x^\prime \,\,\bm{x}^\prime \times 
 \left( \bm{E} (\bm{x}^\prime)
 \times \bm{B} (\bm{x}^\prime) \right) .
\end{equation}
%
By using the standard transverse-longitudinal decomposition of
the vector potential, $\bm{A} = \bm{A}_\perp + \bm{A}_\parallel$, the
electric field can also be decomposed into the transverse and longitudinal
components as
\begin{equation}
 \bm{E} \ = \ \bm{E}_\perp \ + \ \bm{E}_\parallel ,
\end{equation}
where
\begin{equation}
 \bm{E}_\perp \ = \ - \,\frac{\partial \bm{A}_\perp}{\partial t}, \hspace{6mm}
 \bm{E}_\parallel \ = \ - \,\nabla A^0 \ - \ 
 \frac{\partial \bm{A}_\parallel}{\partial t} .
\end{equation}
On the other hand, the magnetic field has the transverse component only, i.e.
we have $\bm{B} = \bm{B}_\perp$, since $\bm{B}_\parallel = \nabla \times
\bm{A}_\parallel = 0$ by definition. The total angular momentum of the
electromagnetic field can therefore be decomposed into two pieces 
as \cite{Waka2010}
\begin{equation}
 \bm{J}^\gamma \ = \ \bm{J}^\gamma_\perp \ + \ \bm{J}^\gamma_\parallel ,
\end{equation}
where
\begin{eqnarray}
 \bm{J}^\gamma_\perp \ &=& \ \int \,d^3 x^\prime \,\,\bm{x}^\prime \times \left(
 \bm{E}_\perp (\bm{x}^\prime) \times \bm{B} (\bm{x}^\prime) \right) , \\
 \bm{J}^\gamma_\parallel \ &=& \ \int \,d^3 x^\prime \,\,\bm{x}^\prime \times \left(
 \bm{E}_\parallel (\bm{x}^\prime) \times \bm{B} (\bm{x}^\prime) \right) .
 \end{eqnarray}
First, after partial integration, the transverse part $\bm{J}^\gamma_\perp$ can be 
transformed into the form \cite{WKZZ2018} : 
\begin{eqnarray}
 \bm{J}^\gamma_\perp \ &=& \ \int \,d^3 x^\prime \,\,
 \bm{E}_\perp (\bm{x}^\prime) \times \bm{A}_\perp (\bm{x}^\prime) \nonumber \\
 &+& \int \,d^3 x^\prime \,\,E^j_\perp (\bm{x}^\prime) \,
 (\bm{x}^\prime \times \nabla^\prime) \,
 A^j_\perp (\bm{x}^\prime) \nonumber \\
 &-& \int \,d^3 x^\prime \,\,\nabla^{\prime j} \,
 \left[ E^j_\perp (\bm{x}^\prime) \,
 (\bm{x}^\prime \times \bm{A}_\perp (\bm{x}^\prime)) \right] .
\end{eqnarray}
The first two terms in $\bm{J}^\gamma_\perp$ respectively correspond to 
spin and orbital angular momentum of the photon, while the third term is
the surface integral term. It is important to recognize that these three 
terms survive even in the {\it free photon limit}, i.e. 
even in the absence of the charged particle
sources for the photon field. In this free photon limit, one confirms
that the above decomposition just corresponds to Leader's relation
(\ref{Eq:Surface_Leader}) for $\bm{J}_{poyn}$.

For an interacting system of photons and charged particles, however,
the longitudinal part of $\bm{J}^\gamma$ also gives nonzero
contributions. With use of partial integration supplemented with 
the Gauss law $\nabla^\prime \cdot \bm{E}_\parallel (\bm{x}^\prime) = 
\rho (\bm{x}^\prime)$, the longitudinal part $\bm{J}^\gamma_\parallel$
can be transformed into the following form \cite{WKZZ2018} : 
\begin{eqnarray}
 \bm{J}^\gamma_\parallel &=& 
 \int \,d^3 x^\prime \,\,\rho \,(\bm{x}^\prime \times \bm{A}_\perp (\bm{x}^\prime))
 \nonumber \\
 &+& \int \,d^3 x^\prime \,\,\nabla^{\prime j} \,
 \left[ (\nabla^{\prime j} \,A^0 (\bm{x}^\prime)) \,
 (\bm{x}^\prime \times \bm{A}_\perp (\bm{x}^\prime)) \right] \nonumber \\
 &-& \int \,\,d^3 x^\prime \,\,\nabla^{\prime j} \,\left[ A^0 (\bm{x}^\prime) \,
 (\bm{x}^\prime \times \nabla^\prime) \,A^j_\perp (\bm{x}^\prime) \right] 
 \nonumber \\
 &-& \int \,d^3 x^\prime \,\, \nabla^\prime \times
 ( A^0 (\bm{x}^\prime) \,\bm{A}_\perp (\bm{x}^\prime)) .
\end{eqnarray}
Here, aside from the first term, the remaining three terms are all surface
integral terms. By using the expression for the charge density
$\rho (\bm{x}^\prime) = q \,\delta^3 (\bm{x}^\prime - \bm{x})$ with
$\bm{x}$ representing the position of a charged particle with the charge $q$,
the first term of the above decomposition can also be expressed as
\begin{equation}
 \int \,d^3 x^\prime \,\,\rho (\bm{x}^\prime) \,
 (\bm{x}^\prime \times \bm{A}_\perp (\bm{x}^\prime))
 \ = \ q \,(\bm{x} \times \bm{A}_\perp (\bm{x}) ).
\end{equation}
This is nothing but the {\it potential angular momentum} $\bm{L}_{pot}$
introduced in \cite{Waka2010},\cite{Waka2011}, 
which has a meaning of angular momentum
stored in the photon field in the presence of the charged particle sources.
Note that the other three surface integral terms all contain
the scalar potential $A^0$. Since $\rho = 0$ and $A^0 = 0$ in the
absence of charged particles, all the terms in $\bm{J}^\gamma_\parallel$
including the potential angular momentum term vanish, i.e. we have
\begin{equation}
 \bm{J}^\gamma_\parallel \ \rightarrow \ 0 ,
\end{equation}
in the {\it free photon limit}.
This conversely means that, in the presence of the charged particle sources,
the potential angular momentum term $\bm{L}_{pot}$ does not vanish. 
It rather constitutes an important piece
of angular momentum {\it contained} in the Poynting (or Belinfante) version
of the total angular momentum of the photon.

Since the nucleon is a composite system of color-charged quarks 
and gluons, if we follow exactly the same logical steps, we are naturally led 
to the conclusion that the total angular momentum of the gluon in the 
nucleon consists of the following three pieces
\begin{equation}
 \bm{J}^G_{total} \ \equiv \ \bm{J}^G_{poyn} \ = \ 
 \bm{J}^G_{spin} \ + \ \bm{L}^G_{g.i.c.} \ + \ \bm{L}_{pot} ,
\end{equation}
aside from the surface integral terms, which we have already proved not to
contribute to the integrated nucleon spin sum rule. 
Although we do not repeat here the discussion given 
in \cite{Waka2010},\cite{Waka2011}, the presence 
of the potential angular momentum
term can also influence the decomposition of the total angular momentum
of quarks. The potential angular momentum is the key quantity that leads
to the existence of two physically inequivalent decompositions of the nucleon
spin, i.e. the decomposition of the canonical type and that of the mechanical
(or kinetic) type. In any case, it should be clear by now that mere discussion of the 
free photon beam would never unravel this important aspect of the
spin decomposition problem in particle physics.   

The role of surface terms in the decomposition of the nucleon spin at the
density level was recently investigated by Lorc\'{e}, Mantovani, and Pasquini,
through the analysis of the spatial distribution of angular
momentum inside the nucleon \cite{LMP2018}. 
It is a widely-known fact that the canonical
energy momentum tensor (EMT) obtained by using the Noether theorem
is in general neither gauge invariant nor symmetric. Utilizing the
arbitrariness of the Noether current such that one can always add any
current which is conserved by itself, Belinfante and Rosenfeld proposed to
add a so-called super potential term to the definition of both the
EMT and the angular momentum tensor \cite{Belinfante1939},\cite{Belinfante1940},
\cite{Rosenfeld1940} :
\begin{eqnarray}
 T^{\mu \nu}_{Bel} (x) \ &=& \ T^{\mu \nu}_{can} (x) \ + \ 
 \partial_\lambda \, \Sigma^{\lambda \mu \nu} (x), \\
 J^{\mu \alpha \beta}_{Bel} (x) \ &=& \ J^{\mu \alpha \beta}_{can} (x) \nonumber \\
 &+& \partial_\lambda \left[ x^\alpha \,\Sigma^{\lambda \mu \beta} (x) \ - \ 
 x^\beta \,\Sigma^{\lambda \mu \alpha} (x) \right] . \ \ \ \ 
\end{eqnarray}
where
\begin{equation}
 \Sigma^{\lambda \mu \nu} (x) \, = \, \frac{1}{2} \,\left[ S^{\lambda \mu \nu} (x)
 \, + \, S^{\mu \nu \lambda} (x) \, - \, S^{\nu \lambda \mu} (x) \right] .
\end{equation}
Here, $T^{\mu \nu}_{can}$ and $J^{\mu \alpha \beta}_{can}$ respectively stand 
for the canonical EMT and the canonical angular momentum tensor, while
$S^{\lambda \mu \nu} (x)$ is spin density operator that is antisymmetric
with respect to the last two indices $\mu$ and $\nu$, i.e. 
\begin{equation}
 S^{\lambda \nu \mu} (x) \ = \ - \,S^{\lambda \mu \nu} (x).
\end{equation}
This enables them to obtain gauge invariant and symmetric EMT and the
gauge invariant angular momentum tensor with desired symmetries.
It is usually believed that the requirement of symmetric EMT is motivated
by general relativity. Alternatively, one can say that the graviton
couples only to the symmetric part of EMT of matter field.
According to Lorc\'{e} et al, however, there is no reason, from particle
physics perspective, to drop the antisymmetric part from the EMT.
A key observation in their analysis is then as follows.
From the conservation of both the EMT and the
angular momentum tensor, it follows that
the EMT can generally be asymmetric, the antisymmetric part being
given by the divergence of the spin density tensor $S^{\mu \alpha \beta}$,
\begin{equation}
 T^{[\alpha \beta]} (x) \ = \ - \,\partial_\mu \,S^{\mu \alpha \beta} (x),
\end{equation}
where $a^{[\mu} b^{\nu]} = a^\mu \,b^\nu - a^\nu \,b^\mu$.

By some strategical reason, instead of starting from the canonical EMT, 
their analysis starts with the Belinfante-improved form of the quark and gluon
parts of EMTs given by
\begin{eqnarray}
 T^{\mu \nu}_{Bel,q} (x) \ &=& \ \frac{1}{4} \,\bar{\psi} (x) \,\gamma^{\{\mu} \,
 i \,{\overset{\leftrightarrow}{D}}\,^{\nu\}} \,\psi (x) , \\
 T^{\mu \nu}_{Bel,G} (x) \ &=& \ - \,2\,\mbox{Tr} 
 \left[ F^{\mu \lambda} (x) \,F^\nu_\lambda (x) \right] \nonumber \\
 &\,& \ + \ \frac{1}{2} \,
 g^{\mu \nu} \,\mbox{Tr} \left[ F^{\rho \sigma} (x) \,F_{\rho \sigma} (x) \right] ,
 \hspace{6mm}
\end{eqnarray}
and the corresponding angular momentum tensor
\begin{eqnarray}
 J^{\mu \alpha \beta}_{Bel,q} (x) \ &=& \ x^\alpha \,T^{\mu \beta}_{Bel,q} (x)
 \ - \ x^\beta \,T^{\mu \alpha}_{Bel,q} (x), \\
 J^{\mu \alpha \beta}_{Bel,G} (x) \ &=& \ x^\alpha \,T^{\mu \beta}_{Bel,G} (x)
 \ - \ x^\beta \,T^{\mu \alpha}_{Bel,G} (x) , \ \ \ \ 
\end{eqnarray}
where $a^{\{\mu} b^{\nu\}} = a^\mu \,b^\nu + a^\nu \,b^\mu$ and
$\overset{\leftrightarrow}{D} = \frac{1}{2} \,
\left( \overset{\rightarrow}{D} - \overset{\leftarrow}{D} \right)$.

As a first step, they compared the quark part of the Belinfante-improved EMT 
with the EMT in what they call the kinetic form given as
\begin{eqnarray}
 T^{\mu \nu}_{kin,q} (x) \ &=& \ \frac{1}{2} \,\bar{\psi} (x) \,\gamma^\mu \,
 i \,{\overset{\leftrightarrow}{D}}\,^\nu \,\psi (x). 
\end{eqnarray}
They notice that this kinetic EMT is obtained from the Belinfante-improved
EMT by adding an antisymmetric piece as
\begin{equation}
 T^{\mu \nu}_{kin,q} (x) \ = \ T^{\mu \nu}_{Bel,q} (x) \ + \ \frac{1}{2} \,
 T^{[\mu \nu]}_q (x) ,
\end{equation}
where
\begin{equation}
 T^{[\alpha \beta]}_q (x) \ = \ - \,\partial_\lambda \,S^{\lambda \alpha \beta}_q (x), 
\end{equation}
with the definition of the quark spin density operator
\begin{eqnarray}
 S^{\lambda \alpha \beta}_q (x) \ = \ \frac{1}{2} \,
 \epsilon^{\alpha \beta \lambda \sigma} \,
 \bar{\psi} (x) \,\gamma_\sigma \,\gamma_5 \,\psi (x) .
\end{eqnarray}
Similarly, by adding the following super potential term
\begin{eqnarray}
 &-& \partial_\lambda \left[ x^\alpha \,\Sigma^{\lambda \mu \beta}_q (x)
 \, - \, x^\beta \,\Sigma^{\lambda \mu \alpha}_q (x) \right] \nonumber \\
 &=& - \,\frac{1}{2} \,\partial_\lambda 
 \left[ x^\alpha \,S^{\lambda \mu \beta}_q (x) \, - \, 
 x^\beta \,S^{\lambda \mu \alpha}_q (x)
 \right] , \ \ \ 
\end{eqnarray}
to the quark part of the Belinfante-improved angular momentum tensor 
$J^{\mu \alpha \beta}_{Bel,q} (x)$, they obtain the relation
\begin{eqnarray}
 &\,& L^{\mu \alpha \beta}_{kin,q} (x) \ + \ S^{\mu \alpha \beta}_q (x) \ = \ 
 J^{\mu \alpha \beta}_{Bel,q} (x) \nonumber \\
 &\,& \hspace{10mm} - \ \frac{1}{2} \,\partial_\lambda
 \left[ x^\alpha \,S^{\lambda \mu \beta}_q (x) \ - \ 
 x^\beta \,S^{\lambda \mu \alpha}_q (x) \right] . \ \ \ \ 
\end{eqnarray}
Here, the first term on the l.h.s. represents the quark orbital angular momentum
density tensor defined by
\begin{equation}
 L^{\mu \alpha \beta}_{kin,q} (x) \ = \ x^\alpha \,T^{\mu \beta}_{kin,q} (x)
 \ - \ x^\beta \,T^{\mu \alpha}_{kin,q} (x),
\end{equation}
whereas the second term on the l.h.s. is just the quark spin density tensor.
Concerning the gluon part, 
they employ the traditional viewpoint that the total angular momentum
of the gluon cannot be decomposed further into its spin and orbital parts
without conflict with the gauge-invariance and the locality, so that they simply
equate the kinetic and Belinfante-improved tensors for the gluon part as
\begin{equation}
 J^{\mu \alpha \beta}_{kin,G} (x) \ = \ J^{\mu \alpha \beta}_{Bel,G} (x) .
\end{equation}
This immediately leads to their basic relation
\begin{eqnarray}
 &\,& L^{\mu \alpha \beta}_{kin,q} (x) \ + \ S^{\mu \alpha \beta}_q (x) 
 \ + \ J^{\mu \alpha \beta}_{kin,G} (x) \nonumber \\
 \, &=& \, 
 J^{\mu \alpha \beta}_{Bel,q} (x) \ + \ J^{\mu \alpha \beta}_{Bel,G} (x)
 \nonumber \\
 &\,& \hspace{15mm} - \ \frac{1}{2} \,\partial_\lambda
 \left[ x^\alpha \,S^{\lambda \mu \beta}_q (x) \ - \ 
 x^\beta \,S^{\lambda \mu \alpha}_q (x) \right] . \hspace{10mm}  
\end{eqnarray} 
The l.h.s. of the above equation is nothing but the local version of
the Ji decomposition, while the r.h.s. is the decomposition in the
Belinfante version plus a surface term. 
In this way, they are led to their central observation that, 
at the local density level, the difference between the Ji decomposition and the 
Belinfante-improved version is characterized by a surface term, 
which is expressed with the quark spin density or the quark axial 
density that can in principle be related to an observable by means of 
neutrino-nucleon scatterings.

In order to relate the above-mentioned angular momentum densities to
observables, what plays a key role are the following two quantities.
The first is the nucleon matrix elements of the general asymmetric EMT
parametrized by five form factors $A(t), B(t), C(t), D(t)$ and $\bar{C} (t)$ as
\begin{eqnarray}
 &\,& \langle p^\prime, \bm{s}^\prime \,|\, T^{\mu \nu} (0) \,|\,p, \bm{s} \rangle
 \ = \ \bar{u} (p^\prime, \bm{s}^\prime) \,\left[ \frac{P^\mu \,P^\nu}{M} A(t)
 \right. \ \ \ \nonumber \\
 &\,& \hspace{3mm} 
 + \ \frac{P^\mu \,i \,\sigma^{\nu \lambda} \,\Delta_\lambda}{M} \,
 \left( A(t) + B(t) + D(t) \right) \nonumber \\ 
 &\,& \hspace{3mm} + \ 
 \frac{\Delta^\mu \,\Delta^\nu \ - \ g^{\mu \nu} \,\Delta^2}{M} \,C(t)  
 \nonumber \\
 &\,& \hspace{3mm} + \  M \,g^{\mu \nu} \,\bar{C} (t) \nonumber \\
 &\,& \hspace{3mm} + \left. 
 \frac{P^\nu \,i \,\sigma^{\mu \lambda} \,\Delta_\lambda}{4 \,M} \,
 \left( A(t) + B(t) - D(t) \right) \right] u(p, \bm{s}) ,  \ \ \ \ \ \ \ 
\end{eqnarray}
where $M$ is the nucleon mass, $\bm{s}$ and $\bm{s}^\prime$ denote the
rest-frame spin of the initial and final nucleon states, respectively, and
$P = (p^\prime + p)/2, \,\Delta = p^\prime - p, \,t = \Delta^2$.
The second is the nucleon matrix elements of the quark spin operator
$S^{\mu \alpha \beta}_q (0)$ parametrized as
\begin{eqnarray}
 \langle p^\prime, \bm{s}^\prime \,|\, S^{\mu \alpha \beta}_q (0) \,|\,p, \bm{s} \rangle
 \!\!&=&\!\! \frac{1}{2} \,\epsilon^{\mu \alpha \beta \lambda} \nonumber \\
 &\,& \hspace{-35mm} \times \ \bar{u} (p^\prime, \bm{s}^\prime) \,
 \left[ \gamma_\lambda \,\gamma_5 \,G^q_A (t) \ + \ 
 \frac{\Delta_\lambda \,\gamma_5}{2 \,M} \,G^q_P (t) \right] \,
 u (p, \bm{s}) , \hspace{8mm}
\end{eqnarray}
where $G^q_A (t)$ and $G^q_P (t)$ are the axial-vector and induced pseudoscalar
form factors.

As was pointed out in their paper, the EMT form factors $A(t), B(t)$ and $C(t)$ can be 
related to leading-twist generalized parton distributions (GPDs), and they are
in principle be observables.
The form factor $\bar{C} (t)$, which has a relation to the trace of EMT, can
be extracted from $\pi N$ scattering amplitudes and so on. Finally, the remaining
form factor $D(t)$, which is connected with the non-symmetric part of EMT, can be
related to the axial-vector form factor as $D_q (t) = - \,G^q_A (t)$, which is
measurable from neutrino-nucleon scatterings.
Using the above parametrization of the nucleon matrix elements of the EMT
as well as the quark spin operator, Lorc\'{e} et al. analyzed several 
spatial distributions of angular momentum in the Belinfante form and in the kinetic forms.
They are spatial distributions of angular momentum in instant form, the corresponding
2-dimensional distributions in the so-called elastic frame, and the analogous 
distributions in light-front form.
Here, we explain the point of their argument by taking the distributions in 
light-front form as an example. 

They start with the definition of the kinetic OAM distribution of quarks
in four dimensional position space
\begin{equation}
 \langle L^z \rangle (x) \ = \ \epsilon^{3 j k} \,x^j_\perp \,\int \,
 \frac{d^2 \bm{\Delta}_\perp \,d \Delta^+}{(2 \,\pi)^2} \,
 e^{\,i \,\Delta \cdot x} \,\langle T^{+ k} \rangle_{LF} ,
\end{equation}
where
\begin{equation}
 \langle T^{\mu \nu} \rangle_{LF} \ \equiv \ \frac{\langle p^\prime, \bm{s} \,|\,
 T^{\mu \nu} (0) \,|\,p, \bm{s} \rangle}{2 \,\sqrt{p^{\prime +} \,p^+}} .
\end{equation}
The corresponding impact parameter distributions in the light-front formalism
are defined in the Drell-Yan (DY) frame where $\Delta^+ = 0$ and
$\bm{P}_\perp = 0$, which amounts to integrating out the four-dimensional
distributions over the light-front coordinate $x^-$. In this frame,
the dependence on the light-front time $x^+$ also drops out.
Then, with the identification $\bm{x}_\perp \rightarrow \bm{b}_\perp$,
the impact-parameter distributions of kinetic OAM and spin can be
represented as
\begin{eqnarray}
 \langle L^z \rangle (\bm{b}_\perp) \!\!&=&\!\! s^z \,\int \,
 \frac{d^2 \bm{\Delta}_\perp}{(2 \,\pi)^2} \,
 e^{\,- \,i \,\bm{\Delta}_\perp \cdot \bm{b}_\perp} \nonumber \\
 &\,& \hspace{10mm} \times 
 \left[ L(t) \ + \ t \,\frac{d L(t)} {d t} \right]_{t = - \,\bm{\Delta}^2_\perp} , 
 \ \ \ \ \ \\
 \langle S^z \rangle (\bm{b}_\perp) \!\! &=& \!\! \frac{1}{2} \,s^z \,\int \,
 \frac{d^2 \bm{\Delta}_\perp}{(2 \,\pi)^2} \,
 e^{\,- \,i \,\bm{\Delta}_\perp \cdot \bm{b}_\perp} \,
 G_A (- \,\bm{\Delta}^2_\perp) . \ \ \ \ \ 
\end{eqnarray}
Here, $L(t)$ is a combination of the EMT form factors given by
\begin{equation}
 L (t) \ = \ \frac{1}{2} \,\left[ A(t) + B(t) + D(t) \right] ,
\end{equation}
while $G_A (t)$ is the axial-vector form factor.

The corresponding impact-parameter distributions of Belinfante improved total
angular momentum and total divergence (surface term) are given by
\begin{eqnarray}
 \langle J^z_{Bel} \rangle (\bm{b}_\perp) &=& s^z \,\int \,
 \frac{d^2 \Delta_\perp}{(2 \,\pi)^2} \,
 e^{\ - \,i \,\bm{\Delta}_\perp \cdot \bm{b}_\perp} \nonumber \\
 &\,& \hspace{10mm} \times
 \left[ J(t) \ + \ t \,\frac{d J(t)}{d t} \right]_{t = - \,\bm{\Delta}^2_\perp} , 
 \ \ \ \ \ \ \ \\
 \langle M^z \rangle (\bm{b}_\perp) &=& - \,\frac{1}{2} \,s^z \,\int \,
 \frac{d^2 \Delta_\perp}{(2 \,\pi)^2} \,
 e^{\ - \,i \,\bm{\Delta}_\perp \cdot \bm{b}_\perp} \nonumber \\
 &\,& \hspace{18mm} \times
 \left[\, t \,\frac{d G_A(t)}{d t} \right]_{t = - \,\bm{\Delta}^2_\perp} .
\end{eqnarray}
Here $J (t)$ is another combination of EMT for factors given by
\begin{equation}
 J (t) \ = \ \frac{1}{2} \,\left[ A(t) + B(t) \right] .
\end{equation}
Using the relation $J(t) \,- \,L(t) = - \frac{1}{2} \,D(t) = \frac{1}{2} \,G_A (t)$,
one can directly check that the following relation holds
\begin{equation}
 \langle L^z \rangle (\bm{b}_\perp) \ + \ \langle S^z \rangle (\bm{b}_\perp) \ = \ 
 \langle J^z_{Bel} \rangle (\bm{b}_\perp) \ + \ \langle M^z \rangle (\bm{b}_\perp) .
 \label{Eq:OAM_impact_density}
\end{equation}
This gives the relation between the local version of quark part of the
nucleon angular momentum in the Ji decomposition \cite{Ji1997} and 
that of the Belinnfante version supplemented with
the surface density term. 
Now, their main assertion is summarized in the following phrase. 
``While superpotential terms
do not play any role at the level of integrated quantities, it is of crucial 
importance to keep track of them at the level of distributions''.
This would be a reasonable statement in itself. Nonetheless, several comments 
are in order on their main claims.

First, it would certainly be true that the decomposition based on
the Belinfante-improved EMT and that in the kinetic version give
different forms of angular momentum decomposition at the density level.
It is also true that the impact-parameter distributions
$\langle L^z \rangle (\bm{b}_\perp)$, 
$\langle S^z \rangle (\bm{b}_\perp)$,
$\langle J^z_{Bel} \rangle (\bm{b}_\perp)$, and
$\langle M^z \rangle (\bm{b}_\perp)$ are all expressed in terms
of the EMT form factors $A(t), B(t), D(t)$ and/or the axial-vector
form factor $G_A (t) = - \,D (t)$.
However, real observables are not the impact-parameter distributions appearing
in Eq.(\ref{Eq:OAM_impact_density}), but the EMT form factors 
$A(t), B(t)$ and $D(t) = - \,G_A (t)$. 
Any of the impact-parameter distributions appearing in both sides 
of (\ref{Eq:OAM_impact_density}) 
are not direct observables. At the best, they may be predicted within some
theoretical framework like Lattice QCD or some effective models of the nucleon,
and a comparison can be made only among those theoretical predictions.
In particular, although the surface term $\langle M^z \rangle (\bm{b}_\perp)$
is expressed with the axial-vector form factor of the nucleon, we do not
have any means to directly measure this distribution. 
The reason is that there is no external probe which 
couples to this surface density.
This point is vitally different from the case of photon laser physics.
Remember that, in this case, the surface term, which characterizes the
difference between the two versions of photon spin density, can in principle
be probed by measuring the spin angular momentum transfer to
the internal angular momentum of an external photon, as emphasized
by Leader.  

Another remark is the following. 
For obtaining the kinetic decomposition or the Ji decomposition
from the Belinfante-improved form,
which enables the decomposition of the total angular momentum of quarks
into the orbital and spin parts, Lorc\'{e} et al. added a super-potential term, 
which is expressed with the quark spin density tensor. 
According to them, this is motivated by the particle physics
viewpoint that there is no reason to drop the antisymmetric part 
from the EMT tensor, since it has inseparable connection with the spin degrees 
of the particles.
If so, why not include the spin density part of the gluon into the
super potential term ? In fact, the relevant super potential should be
\begin{equation}
 \Sigma^{\lambda \mu \nu}_G (x) \, = \, \frac{1}{2} \,\left[
 S^{\lambda \mu \nu}_G (x) \, + \, S^{\mu \nu \lambda}_G (x) 
 \, - \, S^{\nu \lambda \mu}_G (x) \right],
\end{equation}
where 
\begin{equation}
 S^{\lambda \mu \nu}_G (x) \ = \ - \,2 \,\mbox{Tr} \left[
 F^{\lambda \mu} (x) \,A^\nu (x) \ - \ \left( \mu \leftrightarrow \nu \right) \right] ,
\end{equation}
is the gluon spin density operator. Naturally, it is a widely-recognized fact 
that this expression of the gluon spin density is not gauge-invariant. 
However, a natural standpoint could be that the spin degrees of freedom is
more important than the gauge degrees of freedom, since the former is
physical, while the latter is just redundancy with little physical contents.
Alternatively, one can introduce a gluon spin density tensor, which is formally
gauge-invariant. This can be done by introducing the decomposition of
the gluon field into its physical and pure-gauge components as
$A^\mu = A^\mu_{phys} + A^\mu_{pure}$ as was done in \cite{Waka2010},\cite{Waka2011}.
(We point out that the deep meaning hidden in the concept of the
physical component of the gauge field can most transparently be understood
through the analysis of a solvable quantum mechanical system, i.e.
the Landau problem \cite{WKZ2018}.)
Although these two standpoints can easily be compromised, here 
it is simpler to take the first standpoint, in order not to introduce
unnecessary complexity into the argument. 
Adding the corresponding super-potential term to the gluon part
of the Belinfante-improved angular momentum tensor, we readily obtain
\begin{equation}
 J^{\mu \alpha \beta}_{kin,G} \ = \ J^{\mu \alpha \beta}_{Bel,G} \ - \
 \partial_\lambda \left[ x^\alpha \,\Sigma^{\lambda \mu \beta}_G \ - \ 
 x^\beta \,\Sigma^{\lambda \mu \alpha}_G \right] , 
\end{equation}
with
\begin{eqnarray}
 J^{\mu \alpha \beta}_{kin,G} &=& 
 - \ 2 \ \mbox{Tr} \left[ F^{\mu \alpha} \,A^\beta \ - \ 
 F^{\mu \beta} \,A^\alpha \right] \nonumber \\
 &\,& \ + \ 2 \,\,\mbox{Tr} \left[ F^{\lambda \mu} \,\left( x^\alpha \,\partial^\beta
 \ - \ x^\beta \,\partial^\alpha \right) \,A_\lambda \right] \nonumber \\
 &\,& \ + \ 2 \,\,\mbox{Tr} \left[ (D_\lambda \,F^{\lambda \mu}) \,
 \left( x^\alpha \,A^\beta \ - \ x^\beta \,A^\alpha \right) \right] \ \ \ \ 
 \nonumber \\
 &\,& \ + \ \frac{1}{2} \,\,\mbox{Tr} \,F^2 \,
 \left( x^\alpha \,g^{\mu \beta} \ - \ x^\beta \,g^{\mu \alpha} \right) .
\end{eqnarray}
Here, the 1st and the 2nd term of the above equation just correspond to
the spin and canonical orbital angular momentum tensors of the gluon field.
The 3rd term is nothing but the tensor corresponding to the potential
angular momentum, which survives only for coupled quark-gluon system.
The 4th term, sometimes called the boost term, does not contribute to
the angular momentum density, which is of our concern here.

Combining the above result for the gluon part with that for the quark part, 
we are thus led to the following decomposition of the nucleon angular momentum
tensor,
\begin{eqnarray}
 J^{\mu \alpha \beta}_{nucleon} &=& L^{\mu \alpha \beta}_{kin,q} \ + \
 S^{\mu \alpha \beta}_q \ + \ L^{\mu \alpha \beta}_{kin,G} \ + \ 
 S^{\mu \alpha \beta}_G \ \ \ \ \nonumber \\
 &+& \ \mbox{boost} , 
\end{eqnarray}
where
\begin{eqnarray}
 L^{\mu \alpha \beta}_{kin,q} \ &=& \ \frac{1}{2} \,\bar{\psi} \,\gamma^\mu
 \left( x^\alpha \,i \,\overset{\leftrightarrow}{D}\,^\beta \ - \ 
 x^\beta \,i \,\overset{\leftrightarrow}{D}\,^\alpha \right) \psi , \ \ \ 
\end{eqnarray}
while
\begin{eqnarray}
 L^{\mu \alpha \beta}_{kin,G} \ &=& \ 
 2 \,\mbox{Tr} \left[ F^{\lambda \mu} \,\left( x^\alpha \,\partial^\beta 
 \ - \ x^\beta \,\partial^\alpha \right) \,A_\lambda \right] \nonumber \\
 \ &+& \ 2 \,\mbox{Tr} \left[ \left( D_\lambda \,F^{\lambda \mu} \right) \,
 \left( x^\alpha \,A^\beta \ - \ x^\beta \,A^\alpha \right) \right] 
 \ \ \ \ \nonumber \\
 & = & \ \ L^{\mu \alpha \beta}_{can,G} \ \ + \ \ L^{\mu \alpha \beta}_{pot}, 
 \\
 S^{\mu \alpha \beta}_G \ &=& - \ 2 \,\mbox{Tr} \left[ F^{\mu \alpha} \,A^\beta 
 \ - \ 
 F^{\mu \beta} \,A^\alpha \right] .
\end{eqnarray}
Aside from the gauge problem of the gluon part, which can formally be
eliminated by the introduction of the concept of the {\it physical component} 
of the gauge field, the above is precisely the decomposition which we 
proposed in \cite{Waka2010},\cite{Waka2011} as a natural extension of 
the Ji decomposition.

Naturally, there is a reason why Lorc\'{e} et al. did not introduce the
super-potential term corresponding to the gluon spin density,
aside from the problem of gauge invariance.
At variance with the quark spin density term, which can be related to an
observable, i.e. the axial-vector form factor of the nucleon, we do not have
any means to relate the gluon spin density to a direct observable.
One might think that this fact is connected with the color-gauge-variant nature
of the gluon spin density.
However, this understanding would not be necessarily correct.
In fact, suppose that we live in a world where there exists only the
electromagnetic interaction besides the strong interaction, i.e.
in a world with no weak interaction. It is clear that,
in the absence of external weak probe, we have no means to experimentally
access the quark spin density even if the quark spin density satisfies
color-gauge invariance. This implies that there is no absolute
connection between the color-gauge invariance of a certain quantity
and its observability. After all, what is vital for observability is the existence
of external probe which couples to the quantity in question.

We emphasize that the decomposition of the total gluon angular
momentum into its spin and orbital parts is a natural operation also from the 
standpoint of perturbative QCD. The longitudinally polarized gluon
distribution as well as its first moment are well-established concepts
within the framework of perturbative QCD, even though they are
usually thought to be theoretical-scheme-dependent observables. 
At any rate, one should keep in mind the fact that various types of
decomposition of the angular momentum inside the nucleon
are related though an addition or subtraction of surface terms, the density 
of which can hardly be verified by direct measurements.
This implies that discussing the
density-level decomposition of the nucleon spin is a far more
difficult task than discussing the integrated nucleon spin
sum rule, especially if we consider its direct experimental verification seriously.

\section{Conclusion}
\label{sec:5}

When discussing the decomposition of the nucleon spin into its constituents,
it is customary to simply discard the contributions of surface terms.
However, several authors issued a warning against silent neglect of
the surface terms in the nucleon spin decomposition problem, especially
in view of the nontrivial topological configuration of the gluon field in the 
QCD vacuum, which dictates that the gluon field does not vanish 
at the spatial infinity.
In the present paper, we have carefully investigated the role of surface
terms in the nucleon spin decomposition problem.
First, we showed that the surface terms do not contribute to the integrated
sum rule of the nucleon spin, in contradiction with Lowdon's claim that
the surface terms just cancel the contributions of the quark spin and the
gluon spin in the sum rule. The origin of the discrepancy seems to be
the non-commuting nature of the coordinate and momentum operators,
which was overlooked in Lowdon's treatment of the problem.

Also addressed by the present investigation is the question
whether the non-trivial topological configuration of the gluon field in fact
brings about a delta-function-type singularity at the zero Bjorken variable into
the longitudinally polarized gluon distribution, as some authors
claim. This is an important question,
because the existence of such a singularity would spoil the practical
significance of the nucleon spin sum rule. 
As we have explained in full detail, the answer to this question is intricately 
connected with the rigorous and unambiguous definition of
the longitudinally polarized gluon distribution function, which has been left
in obscure status for a long time. After discussing what should be the
most reasonable definition of the longitudinally polarized gluon distribution
from the physical viewpoint, we showed that this fundamental distribution 
does not have a delta-function-type
singularity, thereby supporting the existence of a physically meaningful
decomposition of the nucleon spin.

We have also critically review the recent analysis by Lorc\'{e} et al.,
which explored the role of surface terms on the nucleon angular 
momentum decomposition at the density level. 
They compared the two angular momentum decompositions of the nucleon
at the density level. The one is the so-called
Belinfante-improved version and the other is the widely-known Ji decomposition.
They argue that the Ji decomposition is obtained from the Belinfante 
one by adding a super-potential term, which is expressed with
the quark spin density tensor. According to them, this addition
of the super-potential term is strongly motivated by the particle physics 
perspective that there is no reason to drop the antisymmetric part from the
energy-momentum tensor since it has close connection with the
spin degrees of freedom of the particle. It is important to recognize
the fact that this procedure of adding a super potential term works to 
decompose the total quark angular momentum into its spin and orbital parts. 
In our opinion, however, if one pushes forward this particle physics standpoint 
further, it is natural to include also the spin density part of 
the gluon into the super potential term. 
As expected, this enables us to decompose the total gluon angular momentum
into its spin and orbital parts. 
Somewhat nontrivial is that what is contained in this decomposition are
not only the gluon spin and canonical orbital angular momentum but 
also the what we call the {\it potential angular momentum}. 
The appearance of the potential angular momentum term
is an inevitable consequence of the fact that the gluon in the nucleon is
a bound particle not a free particle. 
The resultant decomposition is nothing but the mechanical
decomposition of the nucleon spin proposed in \cite{Waka2010},\cite{Waka2011}
as a natural extension of the Ji decomposition \cite{Ji1997}. 
Thus, one may be able to say that, from the particle
physics perspective, in which the spin degrees of freedom is thought to be
more important than the unphysical gauge degrees of freedom, 
the Ji decomposition is only a halfway decomposition. 
The full mechanical decomposition, which handles the
spin degrees of freedom of quarks and gluons on the equal footing, is a more 
complete decomposition of the nucleon spin in particle physics.  
Note that this complete decomposition of the nucleon spin is welcome as well as
mandatory also from the standpoint of perturbative QCD, where the gluon spin,
or more generally, the longitudinally polarized gluon distribution is
well-defined and plays an unreplaceable role.

\vspace{3mm}
\section*{\bf Acknowledgments}
The author appreciate many helpful discussions with K. Tanaka,
especially on the delicacy of proper definition of the longitudinally polarized 
gluon distributions. 



\end{document}